\documentclass[11pt]{article}
\usepackage{amsmath,amssymb,latexsym,amsthm,mathtools}
\usepackage[all]{xy}
\usepackage{float}
\usepackage{graphicx,epstopdf,url}
\usepackage{amsfonts}
\def\LARGE{\Large}
\def\Large{\large}

\usepackage{appendix}
\usepackage{enumerate}
\usepackage{multirow}

\usepackage[left=2cm,top=2cm,right=2cm,nohead,foot=2cm]{geometry}

\usepackage{xcolor}

\newcommand\keywords[1]{\textbf{Keywords:} #1}

\begin{document}

\title{An Exploration of Modeling Approaches for Capturing Seasonal Transmission in Stochastic Epidemic Models}
\author{Mahmudul Bari Hridoy\\
Department of Mathematics \& Statistics\\
Texas Tech University\\
Lubbock, Texas 79409-1042}

\date{}

\maketitle

\begin{abstract}

Seasonal variations in the incidence of infectious diseases are a well-established phenomenon, driven by factors such as climate changes, social behaviors, and ecological interactions that influence host susceptibility and transmission rates. While seasonality plays a significant role in shaping epidemiological dynamics, it is often overlooked in both empirical and theoretical studies. Incorporating seasonal parameters into mathematical models of infectious diseases is crucial for accurately capturing disease dynamics, enhancing the predictive power of these models, and developing successful control strategies. This paper highlights key modeling approaches for incorporating seasonality into disease transmission, including sinusoidal functions, periodic piecewise linear functions, Fourier series expansions, Gaussian functions, and data-driven methods, accompanied by real-world examples. Additionally, a stochastic Susceptible-Infected-Recovered (SIR) model with seasonal transmission is demonstrated through numerical simulations. Important outcome measures, such as the basic and instantaneous reproduction numbers and the probability of a disease outbreak using branching process approximation of the Markov chain, are also presented to illustrate the impact of seasonality on disease dynamics.
\end{abstract}


\maketitle

\section{Introduction}

\begin{figure}[htp]
\centering
\includegraphics[width=\textwidth]{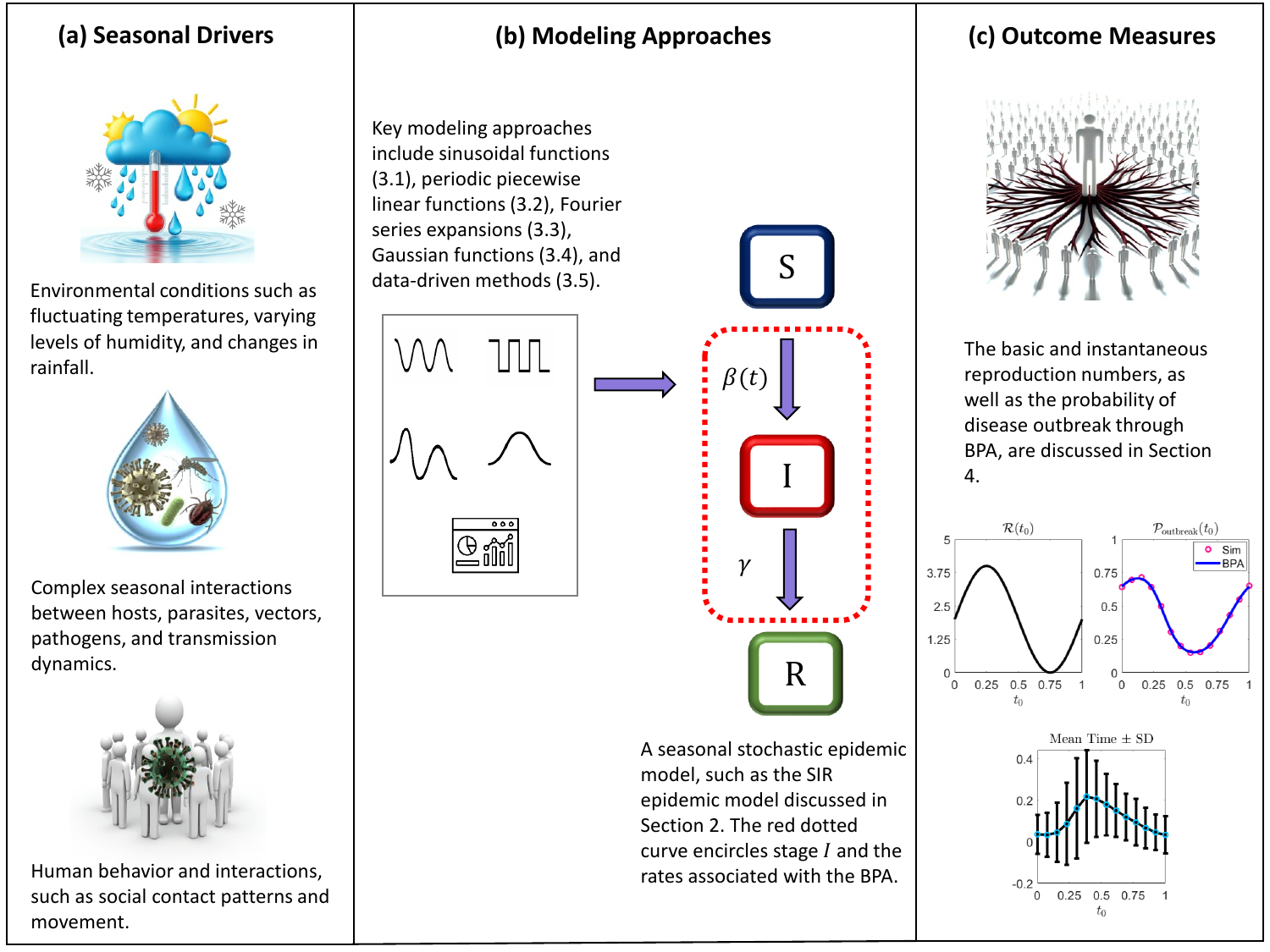} 
\caption{Overview of the seasonal stochastic epidemic model framework, including the compartmental diagram of the seasonal SIR model. The dotted curve encircles stage $I$ and the rates associated with the BPA.}\label{fig:Summary_fig}
\end{figure}

Seasonality is a significant factor influencing the transmission dynamics of infectious diseases, with many pathogens exhibiting distinct seasonal patterns \cite{altizer2006seasonality,fisman2012seasonality,grassly2006seasonal,martinez2018calendar,stone2007seasonal}. Climate change, including shifts in temperature, rainfall, and humidity, further intensifies the complex interactions between hosts, parasites, and vectors \cite{altizer2013climate, buonomo2018seasonality}. These environmental factors, along with social behaviors and ecological interactions, affect host susceptibility and transmission rates, driving seasonal fluctuations in disease prevalence (Figure~\ref{fig:Summary_fig}(a)). This is particularly evident in the transmission of widespread infectious diseases, such as  chickenpox, dengue fever, influenza, Lyme disease, malaria, measles, norovirus, pertussis, rotavirus, and West Nile virus \cite{andreasen2003dynamics, chitnis2012periodically, chowell2008seasonal, coussens2017role, dowell2004seasonality, fine1982measles, hridoy2024synergizing, jutla2013environmental, kronfeld2021drivers, london1973recurrent, nguyen2019modeling, pliego2017seasonality}.

Soper \cite{soper1929interpretation} was arguably the first to address seasonality in epidemic models by incorporating the periodic influx and accumulation of susceptibles, influencing epidemic oscillations based on disease introduction timing and varying susceptibility levels. Subsequent work by Kermack and McKendrick \cite{kermack1927contribution}, and later by Anderson and May \cite{anderson1991populations}, further developed the theory of periodicity in seasonal diseases, emphasizing the role of susceptible dynamics, transmission rates, and demographic factors. Since then, numerous deterministic and stochastic epidemic model studies have explored the nonlinear effects of periodic parameters, particularly their influence on seasonality in disease transmission \cite{aron1984seasonality, bacaer2007approximation, billings2018seasonal, gao2014periodic,nandi2021probability,nipa2020disease, parham2010modeling, prosper2023modeling, suparit2018mathematical, wang2012simple, wang2017dynamics}. 

Several studies have reviewed the impact of seasonality and temporal heterogeneity in contact rates on infectious disease dynamics \cite{buonomo2018seasonality,fisman2012seasonality,grassly2006seasonal,trejos2022dynamics}, providing examples from both human and wildlife systems that highlight the complexities of seasonal environmental drivers \cite{altizer2006seasonality, gage2008climate, sutherst2004global, swei2020patterns}. Other studies have focused on offering an overview of statistical methods for assessing and quantifying the seasonality of infectious diseases \cite{becker1999statistical, christiansen2012methods, madaniyazi2022assessing}, while some have presented mathematical approaches for formulating and simulating stochastic epidemic models \cite{ allen2008introduction,allen2017primer, allen2017predicting,greenwood2009stochastic, Hridoy2024}. Although these studies offer valuable insights into seasonality, heterogeneity, and infectious disease modeling, there remains a need for a detailed examination of modeling approaches specifically tailored to seasonal transmission rates, which this paper aims to address.


Modeling seasonal diseases is challenging, particularly when multiple seasonal drivers are involved. Accurately capturing these effects is crucial for understanding epidemic patterns and developing effective control strategies. This research explores core modeling approaches for incorporating seasonality into disease transmission. These approaches include sinusoidal functions, periodic piecewise linear functions, Fourier series expansions, Gaussian functions, and data-driven methods, all demonstrated through relevant real-world examples. Each approach is assessed in terms of complexity, flexibility, relative advantages, and limitations. Moreover, examples from the literature where these approaches have been applied in epidemic models are cited, and their contributions are briefly discussed.

Additionally, we discuss the application of a seasonal transmission rate in a Susceptible-Infected-Recovered (SIR) model (Figure~\ref{fig:Summary_fig}(b)). Given that epidemic outcomes can vary widely, even with the same initial conditions, we extend this study to a time-nonhomogeneous continuous-time Markov chain (CTMC) SIR model. This extension aims to better understand the combined effects of variability and periodicity on the risk of a disease outbreak, offering a more accurate representation of the unpredictability in disease transmission, especially in small populations and early outbreak phases. Furthermore, important outcome measures such as the basic and instantaneous reproduction numbers, as well as the probabilities of disease extinction and outbreak derived from the branching process approximation (BPA), are briefly reviewed. Numerical simulations are also presented in which the BPA derived from the CTMC SIR model is validated against Monte Carlo simulations to assess the impact of seasonality on disease dynamics.

The paper proceeds as follows: the seasonal SIR epidemic model is discussed in Section~\ref{sec:SIR} and various modeling approaches are presented in Section~\ref{sec:Modeling Approaches}. Outcome measures, including the reproduction numbers, probability of outbreak, and related numerical examples, are discussed in Section~\ref{sec:Outcome Measures}. Finally, Section~\ref{sec:Discussion and Future Directions} provides a summary and discusses future directions for research.

\section{Seasonal SIR epidemic model}
\label{sec:SIR}

In the deterministic SIR model, \(S(t)\), \(I(t)\), and \(R(t)\) represent the number of susceptible, infectious, and recovered individuals, respectively. In the simplest version of the SIR model, only infection and recovery are considered, with a transmission rate of $\beta$ and a recovery rate of $\gamma$, and no births or deaths are included. The seasonal ODE SIR model, with a total population size of $N$ is presented below and the corresponding compartmental diagram is shown in Figure~\ref{fig:Summary_fig}(b).
\begin{equation}\label{eq:SIR_eq}
\left\{
\begin{aligned}
    \frac{dS}{dt} & = -\beta(t) \dfrac{S I}{N}, \\
    \frac{dI}{dt} & = \beta(t) \dfrac{S I}{N} - \gamma I,\\
    \frac{dR}{dt} & = \gamma I. 
\end{aligned}
\right.
\end{equation}


Seasonality is introduced into the transmission rate  $\beta(t)$, which is a positive, time-dependent, periodic function with period $\omega>0$, i.e.,
\begin{equation} \label{eq:periodic_beta}
\beta(t) = \beta(t + \omega), \quad t\in(-\infty,\infty).
\end{equation}

The stochastic SIR model captures the randomness and variability in disease dynamics, particularly when the number of infectious individuals is small or when fluctuations in transmission and other environmental factors significantly affect epidemic outcomes. The CTMC SIR model extends the deterministic framework by using the rates from the ordinary differential equations (ODEs) in Eq.~\eqref{eq:SIR_eq} to define the infinitesimal transition probabilities, with the state variables represented as discrete random variables:

$$
S(t), I(t), R(t) \in \{0,1,2,3,\ldots,N\}, \quad t \in [0,\infty).
$$

The CTMC SIR model is a time-nonhomogeneous Markov process, meaning that the probability of transitioning from one state to another depends on both the current time $t$ and the time elapsed between events. For a small time interval $\Delta t$, the infinitesimal transition probability is defined as:

\begin{equation}
p_{(s,i),(j,k)}(t,t+\Delta t)=P\left(S(t+\Delta t),I(t+\Delta t))= (j,k)|(S(t),I(t)) =(s,i)\right).
\end{equation}

where the transitions occur as follows:
\begin{equation}
p_{(s,i),(j,k)}(t,t+\Delta t) = 
\begin{cases} 
\beta(t) \frac{s i}{N} \Delta t + o(\Delta t) & \text{if } (j,k) = (s-1,i+1) \quad \text{(infection)}, \\
\gamma i \Delta t + o(\Delta t) & \text{if } (j,k) = (s,i-1) \quad \text{(recovery)}, \\
1 - \left( \beta(t)  \frac{s i}{N} + \gamma i \right) \Delta t + o(\Delta t) & \text{if } (j,k) = (s,i) \quad \text{(no change)}, \\
o(\Delta t) & \text{otherwise}.
\end{cases}
\end{equation}

Since the process is nonhomogeneous, the time between events does not follow a simple exponential distribution, making the standard Gillespie algorithm inapplicable \cite{gillespie1977exact}. Instead, a Monte Carlo simulation with small time steps $\Delta t$ is employed to numerically simulate sample paths. 
The events and their corresponding transition rates, which result in changes $\Delta S(t) = S(t+\Delta t) - S(t)$ and $\Delta I(t) = I(t+\Delta t) - I(t)$, are summarized in Table~\ref{tab:sirctmc}.

\begin{table}[H]
\caption{Transitions and rates for the CTMC SIR epidemic model.\label{tab:sirctmc}}
\centering
\begin{tabular}{c|l|l|c}
\hline
{\textbf{Event}} & {\textbf{Description}} & {\textbf{State Transition}} & {\textbf{Rate}}\\ \hline
1 & Infection & $(S,I)\to (S-1,I+1)$ & $\beta(t)S\frac{I}{N}  $ \\ 
2 & Recovery & $(I,R)\to (I-1,R+1)$ & $\gamma I$  \\ \hline
\end{tabular}
\end{table}

\begin{figure}[H]
 \centering
\includegraphics[width=0.7\textwidth]{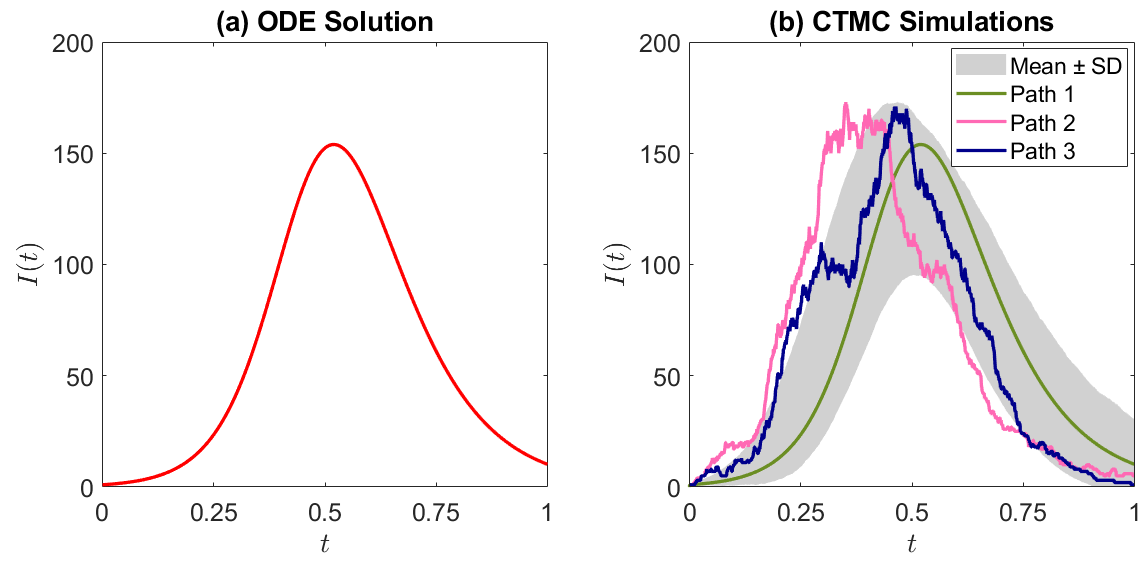} 
\caption{Comparison of the evolution of the number of infected individuals, $I(t)$, over time using both deterministic and stochastic approaches. (a) shows the infected population $I(t)$ from the ODE solution of the seasonal SIR model, while (b) displays three sample paths of the CTMC SIR model, with the shaded area representing the mean $\pm$ one standard deviation across 1000 sample paths.}
\label{fig:sirctmc}
\end{figure}

Both the ODE and CTMC models share the property that $S(t) + I(t) + R(t) = N$, meaning the total population remains constant over time. In both the ODE and CTMC models, when $I(t) = 0$, the system reaches an invariant or absorbing state, indicating that the disease has been eradicated and no further infections can occur, marking the end of the epidemic. Three sample paths for the seasonal CTMC SIR model with $N = 1000$, starting with the initial conditions of a single infected individual, $I(0) = 1$, and $S(0) = 999$, are plotted alongside the solution of the seasonal ODE SIR model in Figure~\ref{fig:sirctmc}.

\section{Modeling approaches}
\label{sec:Modeling Approaches}

This section explores the different modeling approaches for defining $\beta(t)$ in Eq.~\eqref{eq:periodic_beta}, followed by a comparative analysis of the discussed methods.

\subsection{Sinusoidal function}

The sinusoidal function is one of the most commonly used methods for incorporating seasonality into mathematical models. It is widely used in epidemiological studies and public health planning for its straightforward representation. Its use dates back to Joseph Fourier's 1822 work, ``Théorie analytique de la chaleur \cite{baron2003analytical}," which laid the foundation for modeling periodic phenomena. However, its application to infectious disease modeling became more prevalent in the 20th century, particularly in the 1970s for diseases like influenza. Since then, numerous mathematical models have used the sinusoidal function to model seasonal transmission rates \cite{allen2021stochastic, chavez2017sir, huang2018seasonal, husar2024lyme, kamo2002effect, liu2021role, lloyd1996spatial, nipa2021effect}. Generally speaking, a sinusoidal transmission rate can be expressed as follows:
\begin{equation}\label{eq:beta_sinus}
\beta(t) = \bar{\beta} \left(1 + A_{\beta} \cos\left(\frac{2\pi}{\omega}(t + \text{Shift}_{\beta})\right)\right), \quad 0 \leq A_{\beta} \leq 1, \quad 0 \leq Shift_{\beta} < 1.
\end{equation}

where $\beta(t)$ represents the time-varying transmission rate at time $t$, $\bar{\beta}$ is the baseline or average transmission rate around which seasonal fluctuations occur, and $A_{\beta}$ is the amplitude of the sinusoidal function, which determines the extent of seasonal variation. The parameter $\omega$ represents the period of the seasonal cycle (e.g., 12 months for an annual cycle), and $Shift_{\beta}$ is a phase shift parameter that adjusts the timing of the seasonal peak, allowing the model to account for variations in when the peak transmission occurs throughout the year.

One of the major advantages of using a sinusoidal function is its inherent simplicity,  which likely contributes to its widespread use in seasonal modeling. Its straightforward nature makes it relatively easy to incorporate into epidemic models, leading to easier interpretation and implementation. Since sinusoidal functions have a very smooth, regular periodic pattern, they are best suited for seasonal diseases that demonstrate clear, regular seasonal patterns. However, the very regularity of sinusoidal functions limits their flexibility in capturing more irregular, sharp, or abrupt seasonal variations. 

\begin{figure}[H]
 \centering
\includegraphics[width=\textwidth]{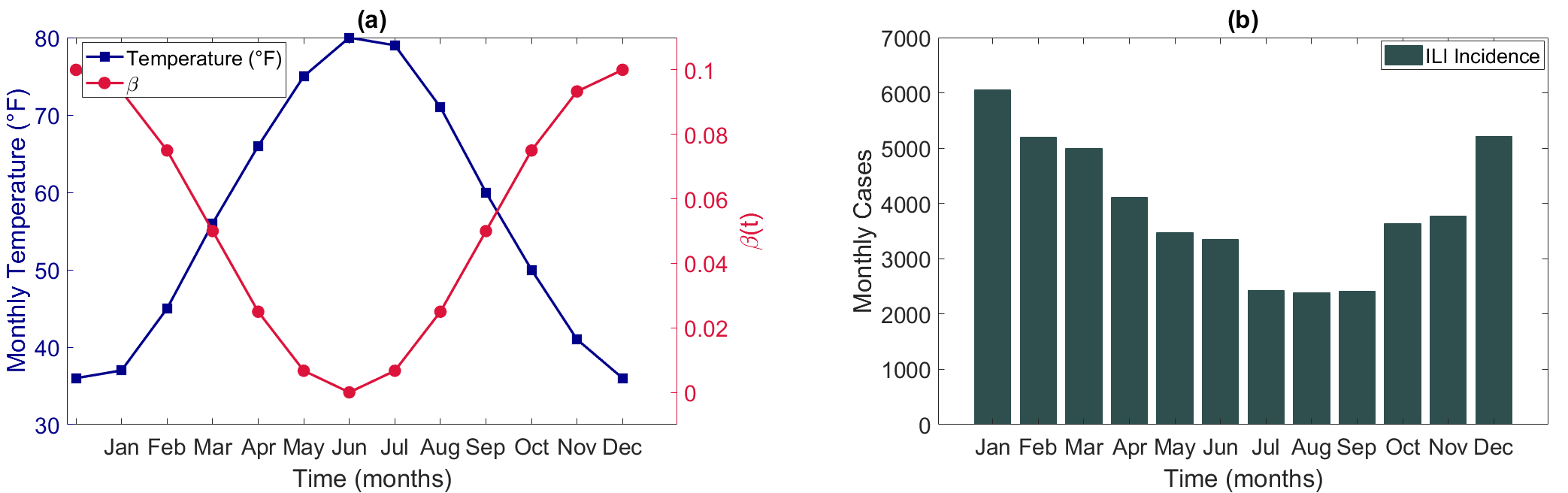} 
\caption{ (a) Time-varying transmission rate $\beta(t)$ and average monthly temperature in New York State over one year. $\beta(t)$ is calculated using Eq.~\eqref{eq:beta_sinus} with the parameters $\bar{\beta} = 0.05$, $A_{\beta} = 1$, $\omega = 12$, and $Shift_{\beta} = -12$. Temperature data for average monthly temperatures in New York State in 2023 is collected from \cite{statista_monthly_temperature_us}. (b) Total number of Influenza-Like Illness (ILI) cases in New York State for the corresponding period in 2023 is sourced from \cite{cdc_fluview_dashboard}
\label{fig:beta_sinus_NYC}}
\end{figure}

Sinusoidal functions are frequently used for diseases where seasonal drivers, such as temperature, are closely linked to transmission. This is evident in diseases like influenza and respiratory syncytial virus (RSV), whose prevalence peaks during colder months. In temperate regions, where flu transmission is significantly influenced by temperature, particularly during the winter,  a sinusoidal transmission rate, as presented in Eq.\eqref{eq:beta_sinus}, is well-suited for modeling the seasonal dynamics of influenza, as demonstrated in \cite{jing2020modeling}. Figure~\ref{fig:beta_sinus_NYC} illustrates the total number of Influenza-Like Illness (ILI) cases in New York State for the year 2023, which can be effectively captured using a sinusoidal transmission rate in conjunction with temperature variations throughout the year.

To overcome the limited flexibility of basic sinusoidal functions, the use of Kot-type functions or similar periodic functions, as proposed in \cite{buonomo2018seasonality}, has gained recent attention in epidemiological modeling. These functions consist of two cosine-modulated terms, making them more complex than simple sinusoidal models, as demonstrated by d’Onofrio et al. in \cite{d2022sir}. The Kot-type function introduces asymmetric weightings of seasonal regimes, allowing it to capture sharper seasonal variations. The Kot-type function used in \cite{d2022sir} is given by:

\begin{equation}\label{eq:beta_kot}
\beta_{\text{Kot}}(t) = \bar{\beta} \left( 1 + \epsilon_\beta \left( \frac{\frac{2}{3} + \cos\left( 2\pi t - Shift_{\beta}\right)}{1 + \frac{2}{3} \cos\left( 2\pi t - Shift_{\beta} \right)} \right) \right)
\end{equation}

where the new parameter $\epsilon_\beta$ controls the strength of the seasonal forcing, determining how strongly the transmission rate fluctuates due to seasonal effects.

\begin{figure}[H]
 \centering
\includegraphics[width=0.5\textwidth]{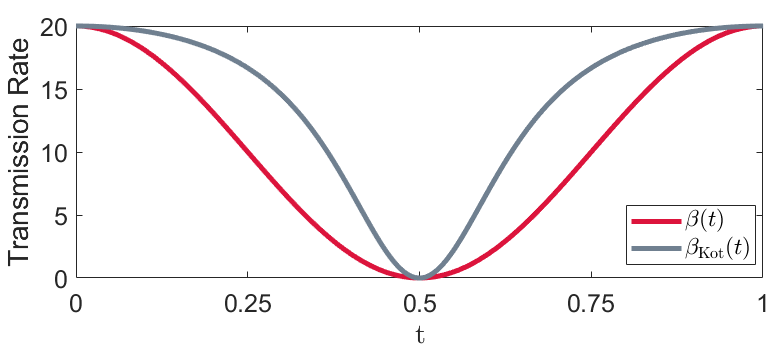} 
\caption{Comparison between the transmission rate obtained from the usual sinusoidal $\beta(t)$ from Eq.~\eqref{eq:beta_sinus} with $\bar{\beta} = 10$, $\text{Shift}_{\beta} = 0$, and $A_{\beta} = 1$, and the transmission rate obtained from the periodic Kot-type function $\beta_{\text{Kot}}(t)$ from Eq.~\eqref{eq:beta_kot} with $\epsilon_\beta = 1$.  All the parameters are scaled per unit year, with time scaled from 0 to 1. 
\label{fig:beta_kot}}
\end{figure}

The term inside the fraction in  Eq.~\eqref{eq:beta_kot}  modulates the periodic behavior of the transmission rate with two cosine terms, allowing for asymmetric seasonal weightings, which makes it more flexible than a simple sinusoidal function. Figure~\ref{fig:beta_kot} shows a comparison between the transmission rate obtained from the usual sinusoidal function and the transmission rate derived from the periodic Kot-type function.

The Kot-type functions allow for sharper seasonal transitions and can capture more complex, non-standard seasonal patterns, particularly in infectious diseases with irregular seasonality or where human behavior significantly impacts transmission cycles (e.g., school terms, vaccination campaigns). These functions provide more control over the shape and timing of seasonal peaks through asymmetrical adjustments, extending their applicability to a wider range of diseases. However, like any complex model, Kot-type functions require accurate seasonal data to calibrate their parameters effectively. Despite these advantages, standard sinusoidal functions remain widely used due to their ease of use and straightforward interpretation.


\subsection{Periodic piecewise linear function}

Instead of using a smooth, continuous function like a sinusoidal curve, the transmission rate can be modeled as a piecewise linear function. This approach accounts for factors such as seasonal weather changes and school terms, dividing the year into distinct periods (e.g., winter, spring, summer, and fall). For example, this could be modeled as:

\begin{equation}\label{eq:beta_piece}
\begin{cases} 
    \beta_1(t) = m_1 t + c_1 & \text{if } t \in \text{Winter} \\
    \beta_2(t) = m_2 t + c_2 & \text{if } t \in \text{Spring} \\
    \beta_3(t) = m_3 t + c_3 & \text{if } t \in \text{Summer} \\
    \beta_4(t) = m_4 t + c_4 & \text{if } t \in \text{Fall}
\end{cases}
\end{equation}

The piece-wise structure of this approach offers better flexibility than sinusoidal functions, allowing it to capture non-smooth seasonal variations and sudden abrupt changes in transmission rates, such as those that might occur during specific periods like school calendars, holidays, or brief intervention periods. However, the sharp spikes, discontinuities, and lack of smoothness in piecewise transmission rates can make the modeling, analysis, and interpretation of seasonal disease dynamics more complex and challenging. Additionally, if calibration with data is required, parameter fitting within each selected time interval can be difficult.

Periodic piecewise linear functions can be an effective approach for modeling seasonal diseases where prevalence shows a sharp, notable spike, with transmission rising or falling dramatically within specific periods. González-Parra et al. \cite{GONZALEZPARRA20093967} employed the Multistage Adomian Decomposition Method (MADM) to solve nonlinear differential equations by dividing the time domain into subdomains, producing a piecewise solution that accurately captures periodic transmission rates, especially in highly variable cases. On the other hand, Silva et al. \cite{silva2022complex} used piecewise constant parameters to model the impact of public health policies and human behavior on the transmission dynamics of COVID-19. Other notable applications which incorporate a periodic piecewise function can be found in \cite{olinky2008seasonal,tanaka2013effects,zhang2019dynamics}. 

\begin{figure}[H]
 \centering
\includegraphics[width=0.65\textwidth]{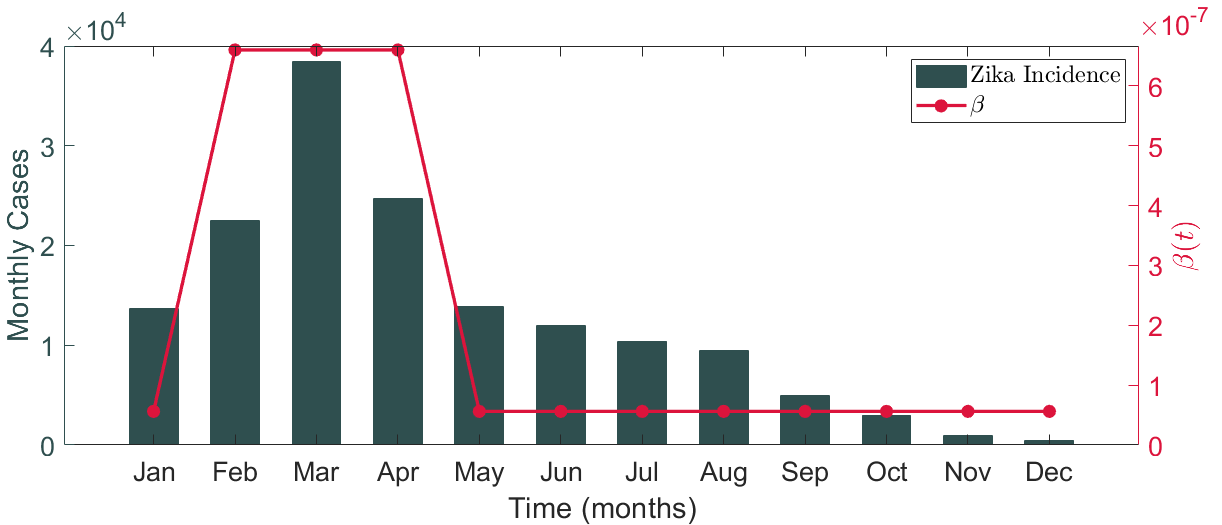} 
\caption{The time-dependent transmission rate is represented as a square wave, as given by Eq.~\eqref{eq:beta_zika}. Zika incidence data from January to December 2016 in Brazil is sourced from \cite{plisa_zika_indicators}.
\label{fig:beta_zika_brasil}}
\end{figure}

Carmona and Gandon \cite{carmona2020winter} used a square wave, which alternates abruptly between two fixed values, to study vector-borne disease emergence and enhance Zika virus risk maps. This method effectively models sharp seasonal changes in mosquito activity, providing a realistic framework for predicting and managing Zika outbreaks by capturing the ‘high’ and ‘low’
transmission states typical of its seasonal dynamics. Figure~\ref{fig:beta_zika_brasil} illustrates the Zika incidence during the 2016 outbreak in Brazil. The transmission rate fluctuates between a baseline level for most of the year and peaks between February and April, indicating a seasonal increase in transmission risk during these months. This fluctuation is represented by the following equation:
\begin{equation}\label{eq:beta_zika}
\beta(t) = 
\begin{cases} 
5.62 \times 10^{-8} & \text{if } t \in \text{Baseline Period (January, May-December)} \\
6.10 \times 10^{-7} & \text{if } t \in \text{Peak Period (February-April)}
\end{cases}
\end{equation}

A natural extension  of periodic piecewise linear functions is the use of piecewise non-linear functions, which offer greater flexibility than linear segments and are better suited to capturing more complex or irregular seasonal patterns. These functions are particularly useful for diseases where transmission is highly dependent on variable factors such as behavior, interventions, or environmental conditions, providing more tailored fits to real-world data. One of the earliest uses was by Capasso and Serio \cite{capasso1978generalization}, who introduced a non-linear incidence rate with a saturated interaction term for modeling cholera transmission. While less common than piecewise linear or sinusoidal functions, piecewise non-linear functions have been used during the COVID-19 pandemic to model transmission changes due to interventions like lockdowns and vaccination rollouts. For instance, Rohith and Devika \cite{rohith2020dynamics} used piecewise non-linear functions to model the nonlinear incidence rate of the COVID-19 pandemic within a Susceptible-Exposed-Infected-Recovered (SEIR) framework. However, while these models offer increased flexibility, it comes at the cost of more complex parameter fitting and interpretation.

\subsection{Fourier series expansion}

In epidemiological modeling, a Fourier series expansion is a useful tool for representing seasonal effects with multiple peaks or varying frequencies. It decomposes the seasonal pattern into a sum of sinusoidal terms, each representing a different harmonic or frequency component. The general form is: 

\begin{equation}\label{eq:beta_fourier}
\beta(t) = \beta_0 + \sum_{n=1}^{N} \left[ a_n \cos\left( \frac{2 \pi n t}{\omega} \right) + b_n \sin\left( \frac{2 \pi n t}{\omega} \right) \right]
\end{equation}

where, $\beta_0$ is the baseline transmission rate, $a_n$ and $b_n$ are the Fourier coefficients for the cosine and sine terms, determining the amplitude of each harmonic, $\omega$ is the period, and $n$ is the harmonic number. 

\begin{figure}[H]
 \centering
\includegraphics[width=0.45\textwidth]{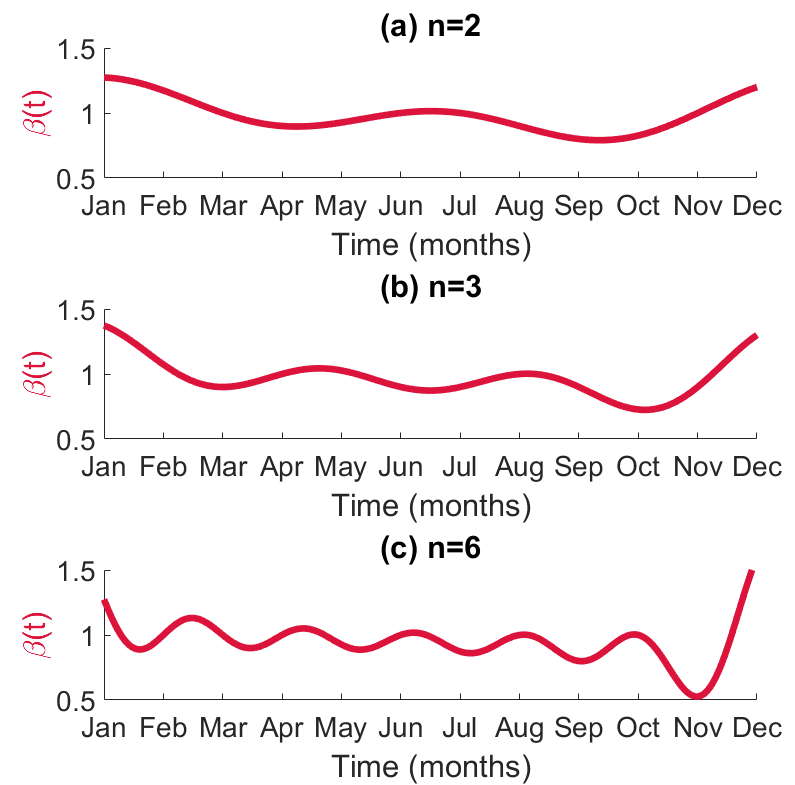} 
\caption{Figure shows $\beta(t)$ with $a_n = b_n = [0.1, 0.1, \dots, 0.1]^\top$ (column vectors of length $n$) and $\omega = 12$ in Eq.~\eqref{eq:beta_fourier}, for the cases of (a) $n=2$, (b) $n=3$, and (c) $n=6$.
\label{fig:beta_fourier}}
\end{figure}

The Fourier series offers versatility in modeling periodic phenomena by adjusting the number of harmonics and coefficients, allowing for the capture of multiple seasonal effects, such as yearly, quarterly, or biannual variations. One key advantage of using a sum of sines and cosines with different frequencies is the ability to capture more complex periodic patterns than a single sinusoidal function, including multi-periodic transmission patterns. For instance, with a small number of harmonics, the series captures simple periodic variations like temperature-driven changes, while a higher $n$ can capture more complex effects, such as multiple peaks due to school vacations or behavioral changes.  Figure~\ref{fig:beta_fourier} shows how different values of $n$ change the number of harmonics in the Fourier series. However, having multiple parameters increases complexity, meaning more parameters to estimate and a higher risk of overfitting.

Fourier series expansions are particularly suitable for seasonal diseases with more complex or multi-periodic transmission patterns, or where multiple seasonal drivers, such as rainfall and temperature, influence the transmission rate. The use of Fourier series as an alternative to sinusoidal models was explored by Gavenčiak et al. \cite{gavenvciak2022seasonal}, enabling an analysis of how results change when applying Fourier series of varying degrees to capture the complex periodic patterns of SARS-CoV-2 seasonality across Europe. Similarly, Deguen et al. \cite{deguen2000estimation} applied Fourier series to model the periodic contact rate in a seasonal SEIR model. The coefficients of the Fourier series were estimated by fitting the model to weekly chickenpox incidence data, providing a more accurate representation of seasonal fluctuations in the contact rate throughout the year. Furthermore, Mortensen et al. \cite{mortensen2024machine} demonstrated how machine learning can enhance predictive accuracy by dynamically adjusting parameters to capture evolving transmission patterns, using a Fourier series to estimate the oscillating infection rate and predict COVID-19 dynamics.

\begin{figure}[H]
 \centering
\includegraphics[width=0.6\textwidth]{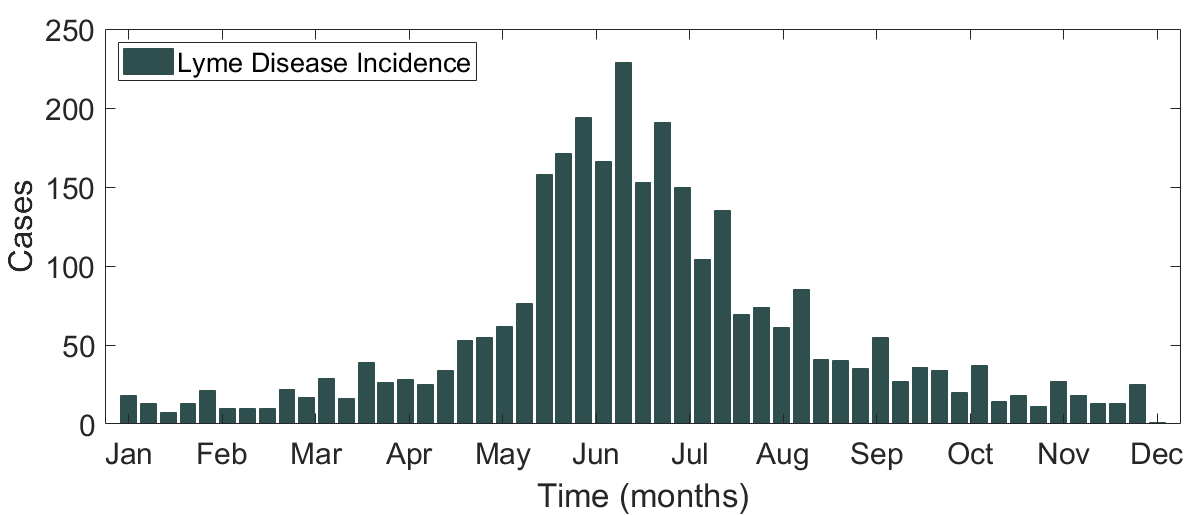} 
\caption{Bars represent weekly Lyme disease cases in the USA for the year 2022, sourced from \cite{CDC_Lyme_2024}.
\label{fig:beta_lyme_USA}}
\end{figure}

Lyme disease transmission involves complex seasonal dynamics, influenced by the activity of ticks and their hosts, which vary over the year. A Fourier series can capture these multiple seasonal peaks and troughs, providing a nuanced model of transmission patterns. In reality, tick activity gradually increases and decreases with changing temperatures and humidity levels, and the Fourier series is well-suited to this type of gradual variation. Figure~\ref{fig:beta_lyme_USA} shows the weekly fluctuations in Lyme disease cases in the USA for 2022, exhibiting a seasonal pattern that justifies using Fourier series to effectively model and capture these periodic trends.

Extensions of Fourier series expansions in epidemic modeling allow for the incorporation of greater complexity to capture intricate seasonal or environmental effects. One such extension involves the use of wavelet transforms, as demonstrated by Kumar et al. in \cite{kumar2020efficient}, where both time and frequency information are captured simultaneously. This method is particularly useful in non-stationary cases where disease transmission patterns change over time. Additionally, the coefficients in the Fourier series expansion can be modeled as functions of time, accommodating evolving seasonality due to interventions, environmental shifts, or behavioral changes, such as changes in vaccine coverage or policy interventions over the years. This makes these approaches more suited to capturing real-world complexities in seasonal disease transmission patterns, especially when patterns are multi-scale or affected by a variety of time-varying factors.


\subsection{Gaussian function}

\begin{figure}[H]
 \centering
\includegraphics[width=0.5\textwidth]{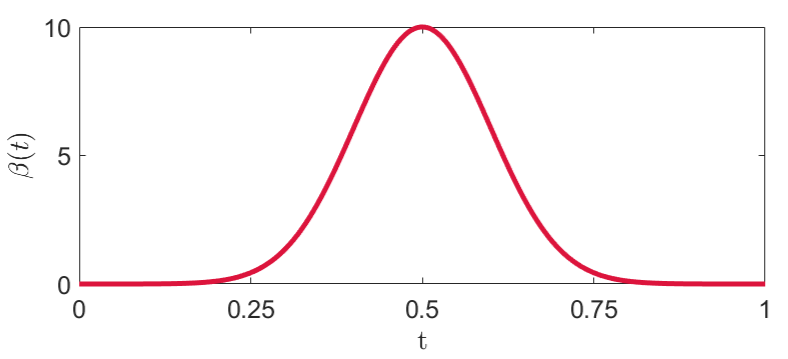} 
\caption{$\beta(t)$ from Eq.~\eqref{eq:beta_gauss}, with parameters set as $A_\beta = 10$, $\mu = 0.5$, and $\sigma = 0.1$. All parameters are scaled per unit year, with time scaled from 0 to 1.
\label{fig:beta_justgauss}}
\end{figure}

A Gaussian function can be effectively used in epidemic models to capture the smooth and symmetrical rise and fall of transmission rates, which often occur during concentrated periods of seasonal transmission in real-world epidemics. Unlike the linear or piecewise functions discussed earlier, which may introduce abrupt changes, the Gaussian function forms a bell-shaped curve (Figure~\ref{fig:beta_justgauss}) that peaks at a specific time (e.g., the peak season for a disease) and gradually decreases before and after the peak. The general form of the Gaussian function is given by:

\begin{equation} \label{eq:beta_gauss}
\beta(t) = A_{\beta} \exp\left( -\frac{(t - \mu)^2}{2\sigma^2} \right)
\end{equation}

where $A_{\beta}$ is the amplitude representing the peak transmission rate, $\mu$ is the mean indicating the peak month of transmission (center of the bell curve), and $\sigma$ is the standard deviation controlling the spread of the curve and determining how quickly transmission rises and falls.

Much like sinusoidal functions, Gaussian functions are relatively easy to implement and interpret, providing a straightforward approach to capturing smooth, natural, well-defined, and concentrated peaks. Additionally, they have low data requirements, with the main parameters of the Gaussian function, the mean and standard deviation, directly mapping to key epidemiological features that drive seasonality. However, unlike more flexible approaches such as Fourier series, the symmetric, predetermined bell shape of the Gaussian function limits its use for diseases with multiple transmission peaks or irregular, asymmetric seasonal patterns.

Gaussian functions are best suited for seasonal diseases where transmission rates rise and fall within a concentrated period, such as during dengue outbreaks with clear seasonal peaks. For instance, Stolerman et al. \cite{stolerman2015sir} used Gaussian-like functions within an SIR-network model to represent the time-dependent transmission rates of dengue fever in Rio de Janeiro, incorporating environmental factors such as rainfall that influence mosquito populations. In like manner, Ramírez-Soto et al. \cite{ramirez2023sir} used a Gaussian function in a SIR-SI model to represent seasonal transmission, allowing the model to capture seasonality observed in dengue outbreaks in Lima, Peru. Several other uses of Gaussian functions in seasonal epidemic models can be found in \cite{alshammari2023analysis,arquam2020impact,setianto2023modeling}.

\begin{figure}[H]
 \centering
\includegraphics[width=\textwidth]{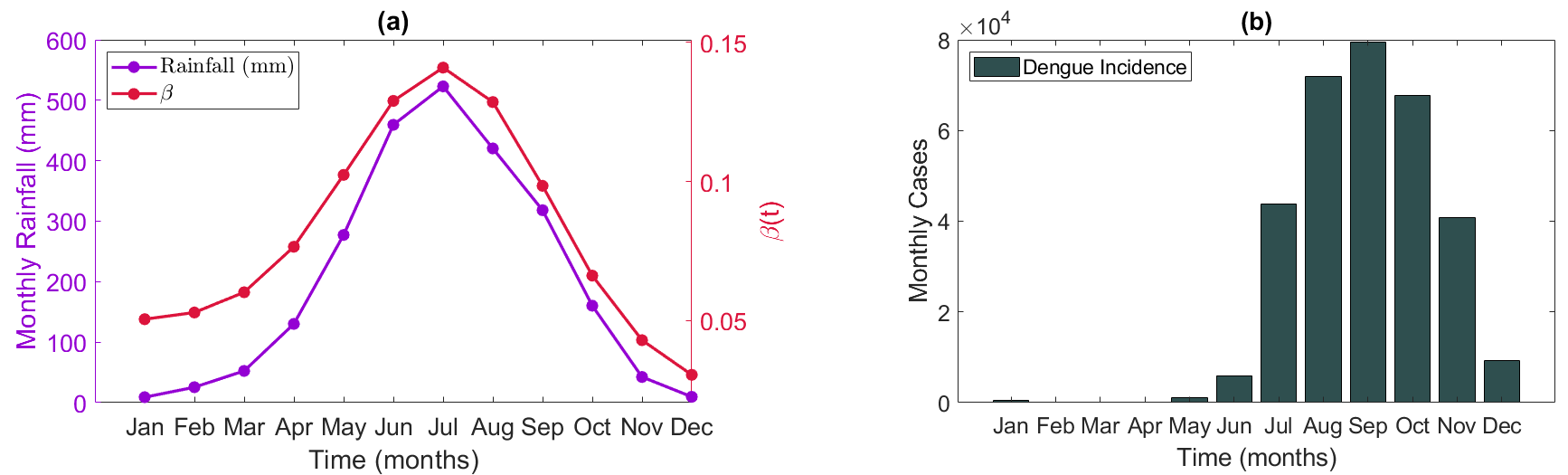} 
\caption{Time-varying transmission rate $\beta(t)$ and average monthly rainfall in Bangladesh over one year. $\beta(t)$ is evaluated using Eq.~\eqref{eq:beta_exp_gauss} with parameters $A_{\beta}=0.10$, $t_{p}=200$, $\sigma=60$, $\beta_{p}=0.03$, and $\beta_{np}=0.02$. Rainfall data for average monthly levels in Bangladesh in 2023 is collected from \cite{bangladesh_meteorological_department_2023}, and Dengue incidence data is sourced from \cite{dghs_dengue_report_2024}.
\label{fig:beta_exp_gauss}}
\end{figure}

To improve the flexibility of the Gaussian function in capturing more complex seasonal transmission patterns, a mixture of multiple Gaussian functions can be used. For example, Setianto et al. \cite{setianto2023modeling} used a Gaussian function to model the time-dependent transmission rate of COVID-19 as a superposition of multiple Gaussian pulses, effectively capturing dynamic changes over time and accurately representing multiple waves of transmission influenced by external factors such as government interventions, behavioral changes, and vaccinations. Additionally, a skewed (or generalized) Gaussian function can be used to model asymmetrical seasonal patterns, or a seasonal modulation factor can be incorporated into the Gaussian function to adjust its amplitude. For example, an extension of the Gaussian function described in Eq.~\eqref{eq:beta_exp_gauss} was used by Hridoy and Mustaquim in \cite{hridoy2024data} to develop an exponentially modulated Gaussian function within a SEIR model, which was employed to capture dengue transmission dynamics in Bangladesh, where the disease is significantly influenced by seasonal variations in rainfall. The functional form of $\beta(t)$ used in \cite{hridoy2024data} is represented as follows:
\begin{equation} \label{eq:beta_exp_gauss}
\beta(t) = \beta_{np} + A_{\beta} \exp\left(-\frac{(t - t_{p})^2}{2\sigma^2}\right) + \frac{\beta_{p}}{1 + \exp\left(\frac{t - t_{p} - \sigma}{\sigma}\right)}
\end{equation}

where $\beta(t)$ is represented as a combination of a baseline transmission rate, $\beta_{np}$ (applicable during non-rainy seasons), and a Gaussian peak transmission rate, $\beta_{p}$ (during rainy seasons), which subsequently decays logistically back to the baseline rate. Here, $A_{\beta}$ denotes the amplitude of seasonal variation, $t_{p}$ indicates the peak time at the midpoint of the rainy season, and $\sigma$ controls the width of the seasonal peak. Figure~\ref{fig:beta_exp_gauss} showcases the exponentially modulated Gaussian function used for modeling the seasonal transmission rate, alongside its correspondence with the average monthly rainfall in Bangladesh for 2023.

\subsection{Data-driven method}

If empirical data is available, a data-driven function can be fitted to reflect the seasonality in the transmission rate. This method directly uses observed data for the transmission rate by integrating real-world factors (e.g., temperature, rainfall, humidity, or vector population changes) instead of relying on a predefined functional form, such as a sinusoidal. The general form of $\beta(t)$ in a data-driven function can be expressed as:

\begin{equation} \label{eq:beta_data}
\beta(t) = f(\text{data}(t), \theta).
\end{equation}

where $f$ is a function that relates the data and parameters to the transmission rate over time, and it can represent a wide range of models, such as machine learning algorithms (e.g., neural networks, random forests), regression models (e.g., generalized additive models), or time-series models (e.g., ARIMA, LSTM); $\text{data}(t)$ represents the set of time-dependent variables that act as seasonal drivers influencing transmission; and $\theta$ is a set of parameters or weights estimated from the data (e.g., regression coefficients, model hyperparameters).

Data-driven methods offer significantly greater flexibility compared to predefined functions, such as those discussed in Eqs.~\eqref{eq:beta_sinus}, \eqref{eq:beta_piece}, \eqref{eq:beta_fourier}, and \eqref{eq:beta_gauss}. These methods can learn patterns directly from empirical data, enabling the capture of complex, non-stationary, irregular, and dynamic seasonal variations, as well as the integration of multiple seasonal drivers. The major advantage of data-driven approaches is their ability to provide superior forecasting performance by learning from historical trends, identifying hidden patterns, and incorporating real-time data. Despite these advantages, data-driven methods typically require large amounts of high-quality data to perform well and may struggle to capture meaningful seasonal patterns if the available data is sparse, noisy, or incomplete. Additionally, machine learning-based data-driven methods can be computationally intensive, prone to overfitting, and difficult to interpret due to the incorporation of multifaceted features and complex interactions \cite{de2015four}. These limitations are not necessarily present in predefined functional forms, such as sinusoidal or piecewise functions, which can perform adequately even with minimal data because they rely on simple assumptions about seasonality.

Data-driven methods can practically be applied to any seasonal disease where data is available, but they are particularly well-suited for diseases with seasonality influenced by multiple factors, such as vector-borne diseases like malaria, dengue, and Zika, where transmission is affected by temperature, rainfall, humidity, and mosquito population dynamics. These methods are also ideal when high-quality data is available, supported by advanced public health surveillance systems or detailed environmental and behavioral datasets. Furthermore, data-driven approaches can be the best choice for fine-tuned forecasting and early detection, as they can leverage real-time data to enable quick action in mitigating outbreaks. This capability has been demonstrated by numerous data-driven models during the COVID-19 pandemic \cite{anastassopoulou2020data, fang2020transmission, kuhl2020data, nabi2020forecasting}.

Several types of data-driven functions can be used to model seasonality or other time-dependent behaviors in epidemiological models. With the abundance of data in the modern world, exciting work has been done in recent times using data-driven seasonal models, e.g., \cite{berry2022seasonality,brooks2015flexible,chintalapudi2020covid, dawa2020seasonal,kuhl2020data,ochieng2024seirs,read2021novel,yuan2021modeling}. An empirical interpolation method, such as linear interpolation, cubic splines, piecewise polynomials, or kernel smoothing, can be used to define $\beta(t)$ based on observed patterns. For example, Prosper et al. \cite{prosper2023modeling} applied cubic spline interpolation to map temperature data to model parameters (e.g., transmission rates, mosquito survival rates), effectively incorporating temperature-dependent traits of mosquitoes and parasites into a malaria transmission model.

Beyond interpolation, other data-driven approaches to model seasonality include regression-based models like Generalized Linear Models (GLMs)  \cite{mcculloch2000generalized}, Generalized Additive Models (GAMs) \cite{hastie2017generalized}, and polynomial regression \cite{ostertagova2012modelling}, which offer flexible alternative for capturing non-linear relationships between transmission rates and seasonal drivers. For instance, Aelenezei et al. \cite{alenezi2021building} used linear and logistic regression within an SIR model to estimate time-dependent parameters, such as transmission, infection, and recovery rates, for COVID-19 transmission in Kuwait, driven primarily by policy measures rather than seasonal factors like temperature. Additional spline methods such as B-splines, P-splines, and thin plate splines can provide smooth, continuous approximations to represent seasonal trends without overfitting the data. 

\begin{figure}[H]
 \centering
\includegraphics[width=0.75\textwidth]{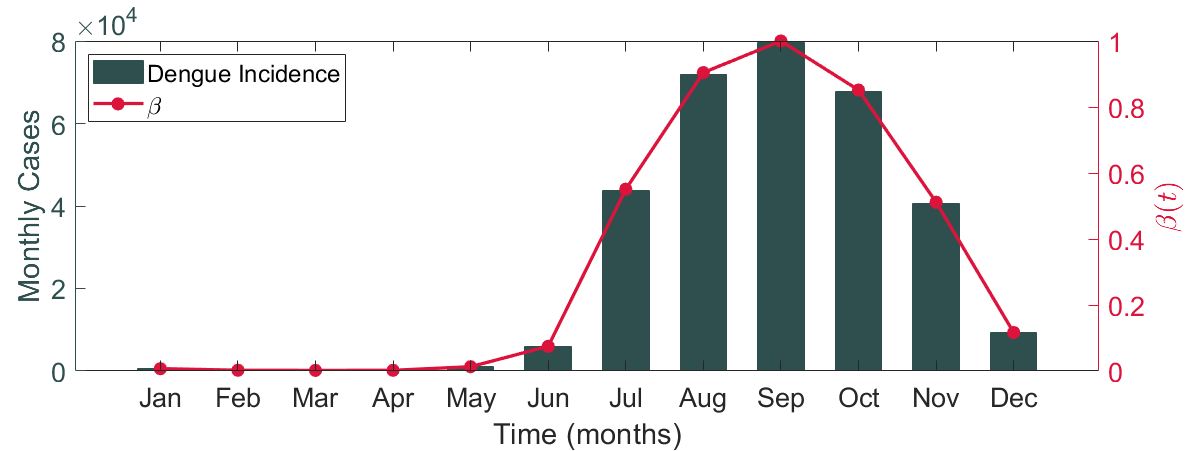} 
\caption{Time-varying transmission rate $\beta(t)$ is evaluated using linear interpolation in conjunction with least-squares optimization to fit the observed dengue incidence data sourced from \cite{dghs_dengue_report_2024}.
\label{fig:beta_data}}
\end{figure}

The accuracy of $\beta(t)$ from Eq.~\eqref{eq:beta_data} is heavily dependent on the availability and reliability of the corresponding data. The choice of the data-driven function significantly influences how well the model captures the actual seasonal variation and smooths the data. Consequently, parameter estimation techniques such as Least Squares Estimation (LSE) \cite{capaldi2012parameter}, Maximum Likelihood Estimation (MLE) \cite{myung2003tutorial}, or Bayesian Estimation \cite{bettencourt2008real} can be complementary methods used to fine-tune data-driven functions for modeling seasonality in epidemiological models, allowing the function’s parameters to be adjusted to best match the empirical data. Figure~\ref{fig:beta_data} shows an example of using linear interpolation in conjunction with least-squares optimization to fit $\beta(t)$ to the observed dengue incidence data from Figure~\ref{fig:beta_exp_gauss} (b).

For further details on an evaluation framework that combines various features, error measures, and ranking schemes for forecast evaluation, see \cite{tabataba2017framework}. For a side-by-side comparison of results from a stochastic agent-based model and a structured metapopulation stochastic model, refer to \cite{ajelli2010comparing}. Additionally, novel concepts in machine learning, such as identifying treatment effects and explaining model outputs, are discussed in \cite{wiemken2020machine}. For a discussion of various machine learning models, hybrid methods that combine mechanistic models with statistical flexibility, and the challenges of deploying these models in real-world forecasting, see \cite{rodriguez2024machine}.


\subsection{Comparison of modeling approaches}

A comparative analysis of various modeling approaches, focusing on key factors such as complexity, data requirements, flexibility, relative advantages and limitations, as well as application examples, is presented in Table~\ref{tab:compare}. The evaluation uses terms like ``Low,'' ``Moderate,'' ``High,'' and ``Very High'' to indicate the relative demands of each method. 

\begin{table}[H]
\caption{Comparative analysis of modeling approaches \label{tab:compare} }
\centering
\small
\begin{tabular}{p{2.5cm}|p{2.5cm}|p{2.5cm}|p{2.5cm}|p{2.5cm}|p{2.5cm}}
\hline
{\textbf{Criteria}}                & {\textbf{Sinusoidal Functions}} & {\textbf{Piecewise Linear Functions}} & {\textbf{Fourier Series Expansions}} & {\textbf{Gaussian Functions}} & {\textbf{Data-Driven Methods}} \\ \hline
{Complexity}              & Low                           & Moderate                           & High                              & Moderate                     & High                         \\ \hline
{Data Requirement}       & Low                      & Low                            & Moderate                          & Low                      & High                         \\ \hline
{Flexibility}             & Low                           & Moderate                           & High                              & Moderate                     & Very High                    \\ \hline
{Suitability} & Seasonal diseases characterized by smooth, predictable, and regular annual cycles & Seasonal diseases with sharp transmission changes within specific periods, characterized by notable spikes    & Seasonal diseases with complex, multi-periodic transmission patterns & Seasonal diseases with a single concentrated peak and unimodal transmission patterns & Seasonal diseases characterized by complex, non-stationary, irregular, and intricate seasonal dynamics\\ \hline
{Advantages}              & Straightforward, easy to implement and interpret & Able to capture abrupt seasonal variations 

& Capable of modeling multiple seasonal components and capturing multiple seasonal peaks & Accurately captures the smooth and symmetrical rise and fall around a well-defined seasonal peak &  Forecasting capabilities, incorporating real-time data, along with the integration of multiple seasonal drivers\\ \hline
{Limitations}             & Assumes regular periodicity, lacks flexibility, and is limited in its ability to capture irregular or abrupt seasonal variations & Discontinuous, requires interval selection, and makes parameter fitting more difficult when calibrating with data & Multiple parameters add complexity to the model, making it harder to estimate and increasing the risk of overfitting & Fails to capture multiple peaks or irregular transmission patterns & Requires large datasets, computationally intensive, and prone to overfitting\\ \hline
{Application Examples}    & \cite{allen2021stochastic,chavez2017sir,jing2020modeling}                  & \cite{carmona2020winter,GONZALEZPARRA20093967,silva2022complex}                   & \cite{deguen2000estimation, gavenvciak2022seasonal,mortensen2024machine}               & \cite{hridoy2024data,ramirez2023sir,stolerman2015sir} & \cite{fang2020transmission,kuhl2020data,prosper2023modeling} \\ \hline
\end{tabular}
\end{table}

\section{Outcome measures}
\label{sec:Outcome Measures}

In this section, we discuss essential outcome measures such as the basic, seasonal, and instantaneous reproduction numbers, as well as the probability of disease outbreak using the branching process approximation from the CTMC SIR epidemic model.

\subsection{Reproduction numbers}

Reproduction numbers are considered one of the most critical metrics for understanding the spread of infectious diseases in epidemic models. Among the different types of reproduction numbers, we will focus on the basic reproduction number ($\mathcal{R}_0$) and the instantaneous reproduction number ($\mathcal{R}(t_0)$). 

For the SIR model, the disease-free equilibrium is given by $S = N$, and $I = R = 0$. The basic reproduction number is simply the ratio of the transmission rate, $\beta$, to the recovery rate, $\gamma$ \cite{diekmann1990definition}, i.e.,
\begin{equation}\label{eq:constR0} 
\mathcal{R}_0 = \frac{\beta}{\gamma}.
\end{equation}

$\mathcal{R}_0$ represents the average number of secondary infections produced by a single infectious individual in a completely susceptible population and can be interpreted as: if $\mathcal{R}_0 > 1$, the disease is expected to spread within the population; if $\mathcal{R}_0 = 1$, the disease will remain stable, neither growing nor declining; and if $\mathcal{R}_0 < 1$, the disease is expected to die out.

For seasonal epidemic models with periodic parameters, the reproduction number differs between a seasonal environment and a constant environment, where all parameters are constant \cite{wang2008threshold}. Wang and Zhao’s approach, which uses Floquet theory \cite{klausmeier2008floquet}, can be applied to calculate the seasonal reproduction number. For models with more complex structures, such as the SEIR model, which includes an additional exposed stage, or other models with multiple infectious stages, the next generation matrix (NGM) is the standard method for computing the basic reproduction number \cite{diekmann2010construction}. However, for simpler models with a single infection stage, such as the SIR model, the basic reproduction number remains unchanged between seasonal and constant environments. The basic reproduction number for the seasonal SIR model with periodic $\beta(t)$, as described in Eq.~\eqref{eq:periodic_beta}, is given by:

\begin{equation}\label{eq:SIRR0}
\mathcal{R}_0 = \frac{\frac{1}{\omega}\displaystyle{\int_0^{\omega}\beta(t)\, dt}}{\gamma} = \frac{\bar{\beta}}{\gamma}.
\end{equation}

To better capture the overall ability of a disease to spread in a periodically changing environment, a time-dependent reproduction number, $\mathcal{R}(t_0)$, referred to as the ``instantaneous reproduction number," is defined as

\begin{equation}\label{eq:instR0} 
\mathcal{R}(t_0) = \frac{\beta(t_0)}{\gamma}, \quad t_0 \in [0, \omega].
\end{equation}

For a comprehensive review on mathematical methods for determining $\mathcal{R}_0$ in ODE disease models and its use in guiding control strategies, see \cite{dietz1993estimation,van2017reproduction}.


\subsection{Probability of disease outbreak}

Although the reproduction numbers computed from deterministic ODE SIR models are important metrics for guiding public health responses to infectious disease outbreaks, the use of a branching
process approximation (BPA) to estimate the probability of disease outbreak is an essential tool for assessing real-world uncertainties that deterministic models cannot capture.

We briefly review the BPA applied to the CTMC SIR epidemic model \cite{allen2017primer, allen2012extinction, nipa2020disease}, with a focus on the infected stage at the disease-free equilibrium (DFE), as depicted within the dotted curve in the compartment diagram in Figure~\ref{fig:Summary_fig}. When the number of infected individuals is small, the BPA assumes negligible interactions between them, where each infected individual independently generates secondary cases.

To compute the seasonal probability of disease extinction (i.e., no major outbreak), the BPA is applied at the DFE, where \(S = N\). Let 
\[
p_{i,j}(t_0, t) = \mathcal{P}(I(t) = j \mid I(t_0) = i), \quad t_0 < t,
\]
represent the probability of transitioning from \(i\) infected individuals at time \(t_0\) to \(j\) infected individuals at time \(t\).

The asymptotic periodic probability of extinction is derived by solving the Backward Kolmogorov Differential Equation (BKDE), applied to the BPA of the CTMC SIR model \cite{allen2017primer}. These backward equations, simpler for computing the probability of extinction than the forward equations, help assess the likelihood of reaching an absorbing state, which marks the end of an epidemic. However, our primary focus is on the stochastic dynamics at the onset of an epidemic, when most of the population remains susceptible. The BKDE for the infected state \(I\) is formulated using the transition probabilities \(p_{i,j}(t_0, t)\):

\[
p_{i,j}(t_0, t) = \beta(\tau)\Delta t_0 \, p_{i+1,j}(t_0 + \Delta t_0, t) + \gamma \Delta t_0 \, p_{i-1,j}(t_0 + \Delta t_0, t)
\]
\[
+ [1 - (\beta(\tau) + \gamma)\Delta t_0] \, p_{i,j}(t_0 + \Delta t_0, t) + o(\Delta t_0),
\]
where \(\tau \in [t_0, t_0 + \Delta t_0]\). By subtracting \(p_{i,j}(t_0 + \Delta t_0, t)\), dividing by \(\Delta t_0\), and taking the limit as \(\Delta t_0 \to 0^+\), we arrive at the BKDE:

\[
-\frac{\partial p_{i,j}(t_0, t)}{\partial t_0} = \beta(t_0) p_{i+1,j}(t_0, t) + \gamma p_{i-1,j}(t_0, t) - (\beta(t_0) + \gamma) p_{i,j}(t_0, t).
\]

Assuming \(i = 1\) and \(j = 0\), with \(p_{1,0} \equiv p_{1,0}(t_0, t)\), and applying the simplifying assumptions \(p_{2,0} = (p_{1,0})^2\) and \(p_{0,0} = 1\), we obtain the BKDE for the probability of extinction:

\[
-\frac{\partial p_{1,0}}{\partial t_0} = \beta(t_0) (p_{1,0})^2 + \gamma - (\beta(t_0) + \gamma) p_{1,0}.
\]

Simplifying further, we get:

\begin{equation}\label{eq:BKDE}
- \frac{\partial p_{1,0}}{\partial t_0} = (\beta(t_0) + \gamma) \left[\frac{\beta(t_0) p_{1,0}^2 + \gamma}{\beta(t_0) + \gamma} - p_{1,0} \right].
\end{equation}

The function inside the square brackets in Eq.~\eqref{eq:BKDE} represents an ``infinitesimal" probability generating function (pgf), also called the ``offspring" pgf \cite{allen2017primer, allen2012extinction}. The asymptotic periodic probability of extinction can be numerically calculated by solving Eq.~\eqref{eq:BKDE} for the BPA of the CTMC SIR model. The probability of extinction, \(P_{\text{extinction}}(1, t_0)\), is given by:

\[
\mathcal{P}_{\text{extinction}}(1, t_0) = \lim_{t \to \infty} p_{1,0}(t_0, t).
\]

The probability of extinction depends on both the number of infected individuals introduced, \(I(t_0) = i\), and the time of introduction, \(t_0\). Due to the independence of infected individuals, the asymptotic probability of disease extinction given \(I(t_0) = i\) is:

\[
\mathcal{P}_{\text{extinction}}(i, t_0) = [\mathcal{P}_{\text{extinction}}(1, t_0)]^i, \quad t_0 \in [0, \omega].
\]

Finally, the probability of a disease outbreak is:

\[
\mathcal{P}_{\text{outbreak}}(i, t_0) = 1 - \mathcal{P}_{\text{extinction}}(i, t_0).
\]

Bacaër and Ait Dads introduced the use of a multitype branching process approximation for predicting seasonal outbreaks in epidemic models and later confirmed that the basic reproduction number (\(\mathcal{R}_0\)) from nonautonomous ODE models with periodic coefficients acts as a threshold for disease extinction \cite{bacaer2014probability}. Similarly, Ball examined the relationship between $\mathcal{R}_0$ and the probability of disease extinction in seasonal epidemic models \cite{ball1983threshold}. When $\mathcal{R}_0 \leq 1$, the probability of disease extinction is one, i.e., $\mathcal{P}_{\text{extinction}}(i,t_0) = 1$. However, when $\mathcal{R}_0 > 1$, the probability of extinction becomes periodic and less than one. This finding implies that seasonality plays a critical role in determining the long-term persistence or extinction of an epidemic, highlighting that an $\mathcal{R}_0 > 1$ is not always sufficient for the disease to establish itself when seasonality is pronounced.

\begin{figure}[H]
 \centering
\includegraphics[width=\textwidth]{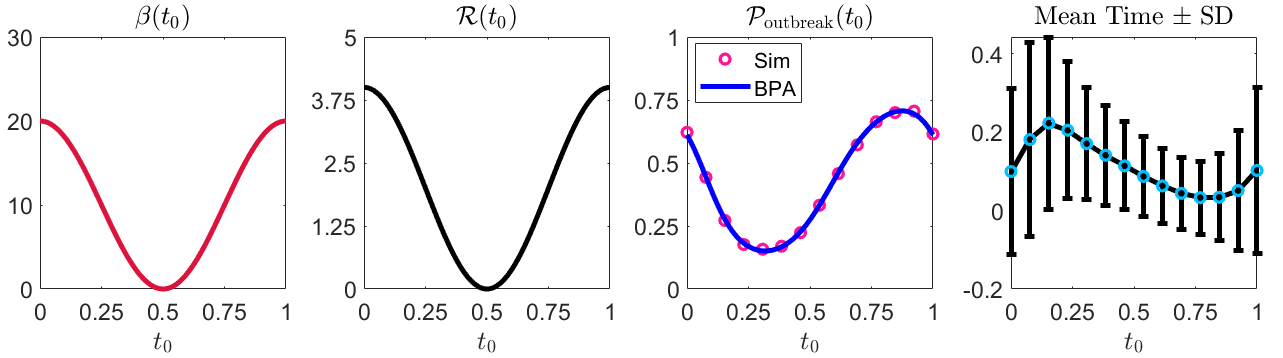} 
\caption{Figure shows $\beta(t_0)$, $\mathcal{R}(t_0)$, $\mathcal{P}_{\text{outbreak}}(t_0)$, and the mean and standard deviation (SD) of the time to extinction, where $\beta(t)$ is modeled using Eq.\eqref{eq:beta_sinus}, with time scaled from 0 to 1 and parameters scaled per unit year. The parameters used are $\bar{\beta} = 10$, $A_{\beta} = 1$, $\omega = 1$, and $Shift_{\beta} = 0$, with a constant recovery rate of $\gamma=\bar{\beta}/2$, resulting in $\mathcal{R}_0=2$. $\mathcal{P}_{\text{outbreak}}(t_0)$ from BPA is verified against the CTMC by simulating $10^4$ sample paths (Sim) and calculating the proportion of infected individuals that reach zero before reaching an outbreak level of 15 individuals, with \(N = 1000\).
\label{fig:pext} }
\end{figure}

Figure~\ref{fig:pext} displays the seasonal transmission rate $\beta(t_0)$, the instantaneous reproduction number $\mathcal{R}(t_0)$, the probability of disease extinction $\mathcal{P}_{\text{extinction}}(t_0)$, as well as the mean and standard deviation (SD) of the time to extinction, where $\beta(t_0)$ is modeled using Eq.\eqref{eq:beta_sinus}.

As shown in Figure~\ref{fig:pext}, the graphs of  $ \beta(t_0) $, $ \mathcal{R}(t_0) $, and $ \mathcal{P}_{\text{outbreak}}(t_0) $ exhibit similar shapes. However, while the extrema of $ \beta(t_0) $ and $ \mathcal{R}(t_0) $ coincide exactly, the extrema of $ \mathcal{P}_{\text{outbreak}}(t_0) $ occur earlier in time. This phenomenon, sometimes referred to as the ``winter is coming'' effect, has been observed across various epidemic models, such as SEIR and vector-host models \cite{carmona2020winter, Hridoy2024, nipa2020disease, nipa2021effect}. Additionally, a Monte Carlo simulation is employed to compute the sample paths of the CTMC models. The results from the simulations show good agreement with the branching process approximation, as illustrated by the simulation circles in Figure~\ref{fig:pext}.

\section{Discussion}
\label{sec:Discussion and Future Directions}

Seasonality plays a pivotal role in the epidemiology of infectious diseases, significantly influencing disease dynamics and informing the development of effective mitigation strategies. This paper provides an overview of various modeling approaches for incorporating seasonality into transmission rates in epidemic models, ranging from simple sinusoidal functions to more advanced data-driven techniques. Each approach offers distinct advantages depending on the complexity of seasonal patterns, the availability of data, and the specific characteristics of the disease under investigation.

Predefined functions, such as sinusoidal and piecewise linear, are popular due to their simplicity and ease of use. These functions perform well when modeling diseases with clear seasonal patterns. However, their simplicity can limit their ability to capture more complex or irregular seasonal trends. Moreover, Gaussian functions can be better suited for modeling single, concentrated peaks, while Fourier series expansions can be more effective at capturing multiple seasonal peaks, making them particularly valuable for diseases with more complex seasonal dynamics.

Extensions that address the limitations of predefined functions have also been discussed in this paper. For instance, the use of Kot-type functions can overcome the restricted control over the shape and timing of seasonal peaks commonly associated with traditional sinusoidal functions, as demonstrated in Figure~\ref{fig:beta_kot}. Similarly, extending a traditional Gaussian function through the use of an exponentially modulated Gaussian function, as shown in Figure~\ref{fig:beta_exp_gauss}(a), can be a valuable enhancement. This extension helps to overcome the limitations of the standard Gaussian function in capturing sharp, narrow peaks, allowing for greater flexibility in representing rapid changes in seasonal patterns.

In contrast to predefined functions, data-driven methods offer greater flexibility as they can adapt to the complexities of real-world disease patterns, particularly when multiple factors influence transmission. Specifically, methods such as empirical interpolation methods, generalized additive models (GAMs), machine learning algorithms, and time-series models excel in capturing intricate seasonal variations driven by multiple variables. These approaches are especially relevant for vector-borne diseases, where transmission is often influenced by interrelated factors, such as environmental changes, host behavior, and vector dynamics.

Although models that incorporate data-driven methods are powerful, they are not without challenges. These approaches rely heavily on large, high-quality datasets. In the absence of reliable data, they may struggle to identify meaningful patterns or risk overfitting to noise, which can reduce their predictive accuracy. Therefore, incorporating parameter estimation techniques such as LSE, MLE, and Bayesian estimation can add significant value to data-driven models, helping to improve their robustness and accuracy. Furthermore, machine learning-based models can be computationally intensive and often lack interpretability, making it difficult to understand the underlying mechanisms driving their predictions. Despite these limitations, the flexibility and adaptability of data-driven methods make them valuable tools in epidemiology, particularly for forecasting and early detection of disease outbreaks.

The comparative analysis of the modeling approaches in Table~\ref{tab:compare} underscores that each method has its own unique strengths and limitations. Ultimately, the selection of an appropriate model depends on several factors, including the specific characteristics of the disease being studied, the quality and availability of data, and the particular research objectives. These considerations are crucial in determining whether a simpler, predefined function or a more flexible, data-driven approach will be most effective in capturing the seasonal dynamics of disease transmission.

Additionally, we discuss the seasonal transmission rate in the context of a seasonal SIR model and extend the deterministic model to a time-nonhomogeneous Markov process, specifically a continuous-time Markov chain (CTMC) SIR model. Key outcome measures, such as the basic reproduction number ($ \mathcal{R}_0 $), the instantaneous reproduction number ($ \mathcal{R}(t_0) $), and the probability of disease outbreak $ \mathcal{P}_{\text{outbreak}}(t_0)$, are also addressed. We derive the probability of an outbreak using the branching process approximation, based on the CTMC SIR model. These tools can provide valuable insights for both modelers and policymakers, aiding in the planning and implementation of effective interventions. As demonstrated by the ``winter is coming'' effect in Figure~\ref{fig:pext}, the extrema of the $ \mathcal{P}_{\text{outbreak}}(t_0) $ occur earlier than those predicted by the seasonal transmission rate, $\beta(t)$. This suggests that disease mitigation strategies or control interventions may need to be implemented in advance of the peaks predicted by the seasonal driver.

While this study focuses on fundamental methods, readers should be aware that other techniques are available, which may also be relevant depending on the modeling context. For instance, predefined functional forms, such as Fourier or Gaussian functions, can be used to define seasonal transmission rates in conjunction with parameters estimated from the data, as demonstrated in \cite{mortensen2024machine} and \cite{hridoy2024data}, respectively. Besides, adding a noise term, as in \cite{otunuga2017global}, can account for environmental variability and random changes (e.g., weather anomalies), or using a hybrid approach that combines a sinusoidal function with step functions \cite{rashidinia2018dynamical} or covariate-based adjustments \cite{tackney2023comparison}, can improve accuracy. 

Although we demonstrate the use of a seasonal transmission rate in a stochastic SIR epidemic model, these approaches are broadly applicable to other epidemic settings, such as SEIR or vector-host models with multiple patches and age structures. Furthermore, beyond modeling seasonality in transmission rates, these approaches can be extended to other seasonal parameters, such as seasonal recovery rates or vector control interventions. Therefore, future research could explore the incorporation of different compartments, structures, and interventions in a stochastic seasonal setting to gain deeper insights into disease dynamics.

On the other hand, incorporating seasonal forcing can lead to more complex disease dynamics, such as multiple stable cycles and chaos \cite{glendinning1997melnikov, he2007epidemiological, nipa2021effectdengue, schwartz1983infinite}. While these complex behaviors are not widely discussed in the literature due to the unpredictability associated with chaos, which limits the practical applicability of such models, the ability to model both stable cycles and chaotic dynamics has significant implications for public health strategies. A promising future direction would be to explore how chaotic dynamics can be harnessed for practical disease forecasting and control. 

Complex interactions between environmental noise, demographic factors, and disease transmission require improved statistical approaches and long-term datasets to fully capture their effects on seasonal disease dynamics. Additionally, human contact patterns remain difficult to model and need further exploration within the context of seasonal epidemics. Future research should prioritize the development of more robust models that integrate both environmental and human behavioral data. A deeper understanding of these factors will enhance the design of control strategies that align with seasonal disease dynamics, leading to more effective interventions.

Given that many infectious diseases are significantly influenced by seasonal patterns, incorporating these variations into epidemiological models is crucial for improving the accuracy of outbreak predictions and peak intensities, as well as for designing timely interventions. Time-heterogeneity, driven by social and environmental factors, remains a highly productive area of research in mathematical and computational epidemiology. Moving forward, it is imperative to prioritize seasonality in infectious disease modeling. As Grassly and Fraser \cite{grassly2006seasonal} aptly noted, ``Seasonality has moved from the center ground of early infectious disease epidemiology to the periphery; a distraction often ignored or assumed of minor importance.'' Re-centering this perspective can lead to improved prediction and prevention of seasonal infectious disease outbreaks.


\section*{Acknowledgments}

The author would like to express sincere gratitude to his Ph.D. advisor, Dr. Angela Peace, for her invaluable guidance and insightful feedback throughout this research. Additionally, the author confirms that no external funding was received to support this study.

\section*{Conflict of interest}

The author declares that there are no conflicts of interest.






\begin{thebibliography}{100}

\bibitem{ajelli2010comparing}
M.~Ajelli, B.~Gon{\c{c}}alves, D.~Balcan, V.~Colizza, H.~Hu, J.~J. Ramasco, S.~Merler, and A.~Vespignani.
\newblock Comparing large-scale computational approaches to epidemic modeling: agent-based versus structured metapopulation models.
\newblock {\em BMC infectious diseases}, 10:1--13, 2010.

\bibitem{alenezi2021building}
M.~N. Alenezi, F.~S. Al-Anzi, and H.~Alabdulrazzaq.
\newblock Building a sensible sir estimation model for covid-19 outspread in kuwait.
\newblock {\em Alexandria Engineering Journal}, 60(3):3161--3175, 2021.

\bibitem{allen2008introduction}
L.~J. Allen.
\newblock An introduction to stochastic epidemic models.
\newblock In {\em Mathematical epidemiology}, pages 81--130. Springer, 2008.

\bibitem{allen2017primer}
L.~J. Allen.
\newblock A primer on stochastic epidemic models: Formulation, numerical simulation, and analysis.
\newblock {\em Infectious Disease Modelling}, 2(2):128--142, 2017.

\bibitem{allen2017predicting}
L.~J. Allen, S.~R. Jang, and L.~I.~T. Roeger.
\newblock Predicting population extinction or disease outbreaks with stochastic models.
\newblock {\em Letters in Biomathematics}, 4(1):1--22, 2017.

\bibitem{allen2012extinction}
L.~J. Allen and G.~E. Lahodny~Jr.
\newblock Extinction thresholds in deterministic and stochastic epidemic models.
\newblock {\em Journal of biological dynamics}, 6(2):590--611, 2012.

\bibitem{allen2021stochastic}
L.~J. Allen and X.~Wang.
\newblock Stochastic models of infectious diseases in a periodic environment with application to cholera epidemics.
\newblock {\em Journal of Mathematical Biology}, 82(6):48, 2021.

\bibitem{alshammari2023analysis}
F.~S. Alshammari.
\newblock Analysis of sirvi model with time dependent coefficients and the effect of vaccination on the transmission rate and covid-19 epidemic waves.
\newblock {\em Infectious Disease Modelling}, 8(1):172--182, 2023.

\bibitem{altizer2006seasonality}
S.~Altizer, A.~Dobson, P.~Hosseini, P.~Hudson, M.~Pascual, and P.~Rohani.
\newblock Seasonality and the dynamics of infectious diseases.
\newblock {\em Ecology Letters}, 9(4):467--484, 2006.

\bibitem{altizer2013climate}
S.~Altizer, R.~S. Ostfeld, P.~T. Johnson, S.~Kutz, and C.~D. Harvell.
\newblock Climate change and infectious diseases: from evidence to a predictive framework.
\newblock {\em science}, 341(6145):514--519, 2013.

\bibitem{anastassopoulou2020data}
C.~Anastassopoulou, L.~Russo, A.~Tsakris, and C.~Siettos.
\newblock Data-based analysis, modelling and forecasting of the covid-19 outbreak.
\newblock {\em PloS one}, 15(3):e0230405, 2020.

\bibitem{anderson1991populations}
R.~M. Anderson.
\newblock Populations and infectious diseases: ecology or epidemiology?
\newblock {\em Journal of Animal Ecology}, 60(1):1--50, 1991.

\bibitem{andreasen2003dynamics}
V.~Andreasen.
\newblock Dynamics of annual influenza a epidemics with immuno-selection.
\newblock {\em Journal of mathematical biology}, 46(6):504--536, 2003.

\bibitem{aron1984seasonality}
J.~L. Aron and I.~B. Schwartz.
\newblock Seasonality and period-doubling bifurcations in an epidemic model.
\newblock {\em Journal of theoretical biology}, 110(4):665--679, 1984.

\bibitem{arquam2020impact}
M.~Arquam, A.~Singh, and H.~Cherifi.
\newblock Impact of seasonal conditions on vector-borne epidemiological dynamics.
\newblock {\em IEEE Access}, 8:94510--94525, 2020.

\bibitem{bacaer2007approximation}
N.~Baca{\"e}r.
\newblock Approximation of the basic reproduction number r 0 for vector-borne diseases with a periodic vector population.
\newblock {\em Bulletin of mathematical biology}, 69:1067--1091, 2007.

\bibitem{bacaer2014probability}
N.~Baca{\"e}r and E.~H. Ait~Dads.
\newblock On the probability of extinction in a periodic environment.
\newblock {\em Journal of Mathematical Biology}, 68(3):533--548, 2014.

\bibitem{ball1983threshold}
F.~Ball.
\newblock The threshold behaviour of epidemic models.
\newblock {\em Journal of Applied Probability}, 20(2):227--241, 1983.

\bibitem{bangladesh_meteorological_department_2023}
{Bangladesh Meteorological Department}.
\newblock {Normal Monthly Rainfall}, 2023.
\newblock [Accessed: August 28, 2024].

\bibitem{baron2003analytical}
J.~B.~J. Baron~Fourier et~al.
\newblock {\em The analytical theory of heat}.
\newblock Courier Corporation, 2003.

\bibitem{becker1999statistical}
N.~G. Becker and T.~Britton.
\newblock Statistical studies of infectious disease incidence.
\newblock {\em Journal of the Royal Statistical Society: Series B (Statistical Methodology)}, 61(2):287--307, 1999.

\bibitem{berry2022seasonality}
I.~Berry, M.~Rahman, M.~S. Flora, T.~Shirin, A.~Alamgir, M.~H. Khan, R.~Anwar, M.~Lisa, F.~Chowdhury, M.~A. Islam, et~al.
\newblock Seasonality of influenza and coseasonality with avian influenza in bangladesh, 2010--19: a retrospective, time-series analysis.
\newblock {\em The Lancet Global Health}, 10(8):e1150--e1158, 2022.

\bibitem{bettencourt2008real}
L.~M. Bettencourt and R.~M. Ribeiro.
\newblock Real time bayesian estimation of the epidemic potential of emerging infectious diseases.
\newblock {\em PloS one}, 3(5):e2185, 2008.

\bibitem{billings2018seasonal}
L.~Billings and E.~Forgoston.
\newblock Seasonal forcing in stochastic epidemiology models.
\newblock {\em Ricerche di Matematica}, 67(1):27--47, 2018.

\bibitem{brooks2015flexible}
L.~C. Brooks, D.~C. Farrow, S.~Hyun, R.~J. Tibshirani, and R.~Rosenfeld.
\newblock Flexible modeling of epidemics with an empirical bayes framework.
\newblock {\em PLoS computational biology}, 11(8):e1004382, 2015.

\bibitem{buonomo2018seasonality}
B.~Buonomo, N.~Chitnis, and A.~d’Onofrio.
\newblock Seasonality in epidemic models: a literature review.
\newblock {\em Ricerche di Matematica}, 67:7--25, 2018.

\bibitem{capaldi2012parameter}
A.~Capaldi, S.~Behrend, B.~Berman, J.~Smith, J.~Wright, and A.~L. Lloyd.
\newblock Parameter estimation and uncertainty quantication for an epidemic model.
\newblock {\em Mathematical biosciences and engineering}, page 553, 2012.

\bibitem{capasso1978generalization}
V.~Capasso and G.~Serio.
\newblock A generalization of the kermack-mckendrick deterministic epidemic model.
\newblock {\em Mathematical biosciences}, 42(1-2):43--61, 1978.

\bibitem{carmona2020winter}
P.~Carmona and S.~Gandon.
\newblock Winter is coming: Pathogen emergence in seasonal environments.
\newblock {\em PLoS computational biology}, 16(7):e1007954, 2020.

\bibitem{CDC_Lyme_2024}
{Centers for Disease Control and Prevention}.
\newblock Lyme disease surveillance data.
\newblock \url{https://www.cdc.gov/lyme/data-research/facts-stats/surveillance-data-1.html}, 2024.
\newblock Accessed: 2024-09-11.

\bibitem{cdc_fluview_dashboard}
{Centers for Disease Control and Prevention (CDC)}.
\newblock Fluview: Influenza surveillance data, 2023.
\newblock Accessed: 2023-09-01.

\bibitem{chavez2017sir}
J.~P. Ch{\'a}vez, T.~G{\"o}tz, S.~Siegmund, and K.~P. Wijaya.
\newblock An sir-dengue transmission model with seasonal effects and impulsive control.
\newblock {\em Mathematical biosciences}, 289:29--39, 2017.

\bibitem{chintalapudi2020covid}
N.~Chintalapudi, G.~Battineni, and F.~Amenta.
\newblock Covid-19 virus outbreak forecasting of registered and recovered cases after sixty day lockdown in italy: A data driven model approach.
\newblock {\em Journal of Microbiology, Immunology and Infection}, 53(3):396--403, 2020.

\bibitem{chitnis2012periodically}
N.~Chitnis, D.~Hardy, and T.~Smith.
\newblock A periodically-forced mathematical model for the seasonal dynamics of malaria in mosquitoes.
\newblock {\em Bulletin of mathematical biology}, 74:1098--1124, 2012.

\bibitem{chowell2008seasonal}
G.~Chowell, M.~Miller, and C.~Viboud.
\newblock Seasonal influenza in the united states, france, and australia: transmission and prospects for control.
\newblock {\em Epidemiology \& Infection}, 136(6):852--864, 2008.

\bibitem{christiansen2012methods}
C.~Christiansen, L.~Pedersen, H.~S{\o}rensen, and K.~Rothman.
\newblock Methods to assess seasonal effects in epidemiological studies of infectious diseases—exemplified by application to the occurrence of meningococcal disease.
\newblock {\em Clinical Microbiology and Infection}, 18(10):963--969, 2012.

\bibitem{coussens2017role}
A.~K. Coussens et~al.
\newblock The role of uv radiation and vitamin d in the seasonality and outcomes of infectious disease.
\newblock {\em Photochemical \& Photobiological Sciences}, 16(3):314--338, 2017.

\bibitem{dawa2020seasonal}
J.~Dawa, G.~O. Emukule, E.~Barasa, M.~A. Widdowson, O.~Anzala, E.~Van~Leeuwen, M.~Baguelin, S.~S. Chaves, and R.~M. Eggo.
\newblock Seasonal influenza vaccination in kenya: an economic evaluation using dynamic transmission modelling.
\newblock {\em BMC medicine}, 18:1--19, 2020.

\bibitem{de2015four}
D.~De~Angelis, A.~M. Presanis, P.~J. Birrell, G.~S. Tomba, and T.~House.
\newblock Four key challenges in infectious disease modelling using data from multiple sources.
\newblock {\em Epidemics}, 10:83--87, 2015.

\bibitem{deguen2000estimation}
S.~Deguen, G.~Thomas, and N.~P. Chau.
\newblock Estimation of the contact rate in a seasonal seir model: application to chickenpox incidence in france.
\newblock {\em Statistics in medicine}, 19(9):1207--1216, 2000.

\bibitem{diekmann2010construction}
O.~Diekmann, J.~Heesterbeek, and M.~G. Roberts.
\newblock The construction of next-generation matrices for compartmental epidemic models.
\newblock {\em Journal of the royal society interface}, 7(47):873--885, 2010.

\bibitem{diekmann1990definition}
O.~Diekmann, J.~A.~P. Heesterbeek, and J.~A.~J. Metz.
\newblock On the definition and the computation of the basic reproduction ratio r 0 in models for infectious diseases in heterogeneous populations.
\newblock {\em Journal of mathematical biology}, 28:365--382, 1990.

\bibitem{dietz1993estimation}
K.~Dietz.
\newblock The estimation of the basic reproduction number for infectious diseases.
\newblock {\em Statistical methods in medical research}, 2(1):23--41, 1993.

\bibitem{dghs_dengue_report_2024}
{Directorate General of Health Services (DGHS), Bangladesh}.
\newblock {Daily Dengue Status Report}, 2024.
\newblock [Accessed: August 28, 2024].

\bibitem{d2022sir}
A.~d'Onofrio, J.~Duarte, C.~Janu{\'a}rio, and N.~Martins.
\newblock A sir forced model with interplays with the external world and periodic internal contact interplays.
\newblock {\em Physics Letters A}, 454:128498, 2022.

\bibitem{dowell2004seasonality}
S.~F. Dowell and M.~S. Ho.
\newblock Seasonality of infectious diseases and severe acute respiratory syndrome--what we don't know can hurt us.
\newblock {\em The Lancet infectious diseases}, 4(11):704--708, 2004.

\bibitem{fang2020transmission}
Y.~Fang, Y.~Nie, and M.~Penny.
\newblock Transmission dynamics of the covid-19 outbreak and effectiveness of government interventions: A data-driven analysis.
\newblock {\em Journal of medical virology}, 92(6):645--659, 2020.

\bibitem{fine1982measles}
P.~E. Fine and J.~A. Clarkson.
\newblock Measles in england and wales—i: an analysis of factors underlying seasonal patterns.
\newblock {\em International journal of epidemiology}, 11(1):5--14, 1982.

\bibitem{fisman2012seasonality}
D.~Fisman.
\newblock Seasonality of viral infections: mechanisms and unknowns.
\newblock {\em Clinical Microbiology and Infection}, 18(10):946--954, 2012.

\bibitem{gage2008climate}
K.~L. Gage, T.~R. Burkot, R.~J. Eisen, and E.~B. Hayes.
\newblock Climate and vectorborne diseases.
\newblock {\em American journal of preventive medicine}, 35(5):436--450, 2008.

\bibitem{gao2014periodic}
D.~Gao, Y.~Lou, and S.~Ruan.
\newblock A periodic ross-macdonald model in a patchy environment.
\newblock {\em Discrete and continuous dynamical systems. Series B}, 19(10):3133, 2014.

\bibitem{gavenvciak2022seasonal}
T.~Gaven{\v{c}}iak, J.~T. Monrad, G.~Leech, M.~Sharma, S.~Mindermann, S.~Bhatt, J.~Brauner, and J.~Kulveit.
\newblock Seasonal variation in sars-cov-2 transmission in temperate climates: A bayesian modelling study in 143 european regions.
\newblock {\em PLoS computational biology}, 18(8):e1010435, 2022.

\bibitem{gillespie1977exact}
D.~T. Gillespie.
\newblock Exact stochastic simulation of coupled chemical reactions.
\newblock {\em The journal of physical chemistry}, 81(25):2340--2361, 1977.

\bibitem{glendinning1997melnikov}
P.~Glendinning and L.~P. Perry.
\newblock Melnikov analysis of chaos in a simple epidemiological model.
\newblock {\em Journal of Mathematical Biology}, 35:359--373, 1997.

\bibitem{GONZALEZPARRA20093967}
G.~González-Parra, A.~J. Arenas, and L.~Jódar.
\newblock Piecewise finite series solutions of seasonal diseases models using multistage adomian method.
\newblock {\em Communications in Nonlinear Science and Numerical Simulation}, 14(11):3967--3977, 2009.

\bibitem{grassly2006seasonal}
N.~C. Grassly and C.~Fraser.
\newblock Seasonal infectious disease epidemiology.
\newblock {\em Proceedings of the Royal Society of London B: Biological Sciences}, 273(1600):2541--2550, 2006.

\bibitem{greenwood2009stochastic}
P.~E. Greenwood and L.~F. Gordillo.
\newblock Stochastic epidemic modeling.
\newblock {\em Mathematical and statistical estimation approaches in epidemiology}, pages 31--52, 2009.

\bibitem{hastie2017generalized}
T.~J. Hastie.
\newblock Generalized additive models.
\newblock In {\em Statistical models in S}, pages 249--307. Routledge, 2017.

\bibitem{he2007epidemiological}
D.~He and D.~J. Earn.
\newblock Epidemiological effects of seasonal oscillations in birth rates.
\newblock {\em Theoretical population biology}, 72(2):274--291, 2007.

\bibitem{plisa_zika_indicators}
{Health Information Platform for the Americas (PLISA)}.
\newblock Zika indicators.
\newblock \url{http://www.paho.org/data/index.php/en/?option=com_content&view=article&id=524&Itemid=}, n.d.
\newblock Accessed: 2019-02-10.

\bibitem{Hridoy2024}
M.~B. Hridoy and L.~J.~S. Allen.
\newblock Investigating seasonal disease emergence and extinction in stochastic epidemic models.
\newblock {\em Mathematical Biosciences}, August 2024.
\newblock Submitted.

\bibitem{hridoy2024data}
M.~B. Hridoy and S.~Mustaquim.
\newblock Data-driven modeling of seasonal dengue dynamics in bangladesh: A bayesian-stochastic approach.
\newblock {\em arXiv preprint arXiv:2410.00947}, 2024.

\bibitem{hridoy2024synergizing}
M.~B. Hridoy and A.~Peace.
\newblock Synergizing health strategies: Exploring the interplay of treatment and vaccination in an age-structured malaria model.
\newblock {\em medRxiv}, pages 2024--09, 2024.

\bibitem{huang2018seasonal}
J.~Huang, S.~Ruan, X.~Wu, and X.~Zhou.
\newblock Seasonal transmission dynamics of measles in china.
\newblock {\em Theory in Biosciences}, 137(2):185--195, 2018.

\bibitem{husar2024lyme}
K.~Husar, D.~C. Pittman, J.~Rajala, F.~Mostafa, and L.~J. Allen.
\newblock Lyme disease models of tick-mouse dynamics with seasonal variation in births, deaths, and tick feeding.
\newblock {\em Bulletin of Mathematical Biology}, 86(3):25, 2024.

\bibitem{jing2020modeling}
S.-L. Jing, H.-F. Huo, and H.~Xiang.
\newblock Modeling the effects of meteorological factors and unreported cases on seasonal influenza outbreaks in gansu province, china.
\newblock {\em Bulletin of Mathematical Biology}, 82:1--36, 2020.

\bibitem{jutla2013environmental}
A.~Jutla, E.~Whitcombe, N.~Hasan, B.~Haley, A.~Akanda, A.~Huq, M.~Alam, R.~B. Sack, and R.~Colwell.
\newblock Environmental factors influencing epidemic cholera.
\newblock {\em The American journal of tropical medicine and hygiene}, 89(3):597, 2013.

\bibitem{kamo2002effect}
M.~Kamo and A.~Sasaki.
\newblock The effect of cross-immunity and seasonal forcing in a multi-strain epidemic model.
\newblock {\em Physica D: Nonlinear Phenomena}, 165(3-4):228--241, 2002.

\bibitem{kermack1927contribution}
W.~O. Kermack and A.~G. McKendrick.
\newblock A contribution to the mathematical theory of epidemics.
\newblock {\em Proceedings of the royal society of london. Series A, Containing papers of a mathematical and physical character}, 115(772):700--721, 1927.

\bibitem{klausmeier2008floquet}
C.~A. Klausmeier.
\newblock Floquet theory: a useful tool for understanding nonequilibrium dynamics.
\newblock {\em Theoretical Ecology}, 1:153--161, 2008.

\bibitem{kronfeld2021drivers}
N.~Kronfeld-Schor, T.~J. Stevenson, S.~Nickbakhsh, E.~S. Schernhammer, X.~C. Dopico, T.~Dayan, M.~Martinez, and B.~Helm.
\newblock Drivers of infectious disease seasonality: potential implications for covid-19.
\newblock {\em Journal of biological rhythms}, 36(1):35--54, 2021.

\bibitem{kuhl2020data}
E.~Kuhl.
\newblock Data-driven modeling of covid-19—lessons learned.
\newblock {\em Extreme Mechanics Letters}, 40:100921, 2020.

\bibitem{kumar2020efficient}
S.~Kumar, A.~Ahmadian, R.~Kumar, D.~Kumar, J.~Singh, D.~Baleanu, and M.~Salimi.
\newblock An efficient numerical method for fractional sir epidemic model of infectious disease by using bernstein wavelets.
\newblock {\em Mathematics}, 8(4):558, 2020.

\bibitem{liu2021role}
X.~Liu, J.~Huang, et~al.
\newblock The role of seasonality in the spread of covid-19 pandemic.
\newblock {\em Environmental Research}, 195:110874, 2021.

\bibitem{lloyd1996spatial}
A.~L. Lloyd and R.~M. May.
\newblock Spatial heterogeneity in epidemic models.
\newblock {\em Journal of theoretical biology}, 179(1):1--11, 1996.

\bibitem{london1973recurrent}
W.~P. London and J.~A. Yorke.
\newblock Recurrent outbreaks of measles, chickenpox and mumps: I. seasonal variation in contact rates.
\newblock {\em American journal of epidemiology}, 98(6):453--468, 1973.

\bibitem{madaniyazi2022assessing}
L.~Madaniyazi, A.~Tobias, Y.~Kim, Y.~Chung, B.~Armstrong, and M.~Hashizume.
\newblock Assessing seasonality and the role of its potential drivers in environmental epidemiology: a tutorial.
\newblock {\em International journal of epidemiology}, 51(5):1677--1686, 2022.

\bibitem{martinez2018calendar}
M.~E. Martinez.
\newblock The calendar of epidemics: Seasonal cycles of infectious diseases.
\newblock {\em PLoS Pathogens}, 14(11):e1007327, 2018.

\bibitem{mcculloch2000generalized}
C.~E. McCulloch.
\newblock Generalized linear models.
\newblock {\em Journal of the American Statistical Association}, 95(452):1320--1324, 2000.

\bibitem{mortensen2024machine}
P.~Mortensen, K.~Lauer, S.~P. Rautenbach, M.~Gallotta, N.~Sharapova, I.~Takkides, M.~Wright, and M.~Linley.
\newblock A machine learning-enabled sir model for adaptive and dynamic forecasting of covid-19.
\newblock {\em medRxiv}, pages 2024--07, 2024.

\bibitem{myung2003tutorial}
I.~J. Myung.
\newblock Tutorial on maximum likelihood estimation.
\newblock {\em Journal of mathematical Psychology}, 47(1):90--100, 2003.

\bibitem{nabi2020forecasting}
K.~N. Nabi.
\newblock Forecasting covid-19 pandemic: A data-driven analysis.
\newblock {\em Chaos, Solitons \& Fractals}, 139:110046, 2020.

\bibitem{nandi2021probability}
A.~Nandi and L.~J. Allen.
\newblock Probability of a zoonotic spillover with seasonal variation.
\newblock {\em Infectious Disease Modelling}, 6:514--531, 2021.

\bibitem{nguyen2019modeling}
A.~Nguyen, J.~Mahaffy, and N.~K. Vaidya.
\newblock Modeling transmission dynamics of lyme disease: Multiple vectors, seasonality, and vector mobility.
\newblock {\em Infectious Disease Modelling}, 4:28--43, 2019.

\bibitem{nipa2020disease}
K.~F. Nipa and L.~J. Allen.
\newblock Disease emergence in multi-patch stochastic epidemic models with demographic and seasonal variability.
\newblock {\em Bulletin of Mathematical Biology}, 82(12):152, 2020.

\bibitem{nipa2021effect}
K.~F. Nipa and L.~J. Allen.
\newblock The effect of demographic variability and periodic fluctuations on disease outbreaks in a vector--host epidemic model.
\newblock {\em Infectious Diseases and Our Planet}, pages 15--35, 2021.

\bibitem{nipa2021effectdengue}
K.~F. Nipa, S.~R.-J. Jang, and L.~J. Allen.
\newblock The effect of demographic and environmental variability on disease outbreak for a dengue model with a seasonally varying vector population.
\newblock {\em Mathematical Biosciences}, 331:108516, 2021.

\bibitem{ochieng2024seirs}
F.~O. Ochieng.
\newblock Seirs model for malaria transmission dynamics incorporating seasonality and awareness campaign.
\newblock {\em Infectious Disease Modelling}, 9(1):84--102, 2024.

\bibitem{olinky2008seasonal}
R.~Olinky, A.~Huppert, and L.~Stone.
\newblock Seasonal dynamics and thresholds governing recurrent epidemics.
\newblock {\em Journal of mathematical biology}, 56:827--839, 2008.

\bibitem{ostertagova2012modelling}
E.~Ostertagov{\'a}.
\newblock Modelling using polynomial regression.
\newblock {\em Procedia engineering}, 48:500--506, 2012.

\bibitem{otunuga2017global}
O.~M. Otunuga.
\newblock Global stability of nonlinear stochastic sei epidemic model with fluctuations in transmission rate of disease.
\newblock {\em International Journal of Stochastic Analysis}, 2017(1):6313620, 2017.

\bibitem{parham2010modeling}
P.~E. Parham and E.~Michael.
\newblock Modeling the effects of weather and climate change on malaria transmission.
\newblock {\em Environmental health perspectives}, 118(5):620--626, 2010.

\bibitem{pliego2017seasonality}
E.~P. Pliego, J.~Vel{\'a}zquez-Castro, and A.~F. Collar.
\newblock Seasonality on the life cycle of aedes aegypti mosquito and its statistical relation with dengue outbreaks.
\newblock {\em Applied Mathematical Modelling}, 50:484--496, 2017.

\bibitem{prosper2023modeling}
O.~Prosper, K.~Gurski, M.~Teboh-Ewungkem, A.~Peace, Z.~Feng, M.~Reynolds, and C.~Manore.
\newblock Modeling seasonal malaria transmission.
\newblock {\em Letters in Biomathematics}, 10(1):3--27, 2023.

\bibitem{ramirez2023sir}
M.~C. Ram{\'\i}rez-Soto, J.~V.~B. Machuca, D.~H. Stalder, D.~Champin, M.~G. M{\'a}rtinez-Fern{\'a}ndez, and C.~E. Schaerer.
\newblock Sir-si model with a gaussian transmission rate: Understanding the dynamics of dengue outbreaks in lima, peru.
\newblock {\em Plos one}, 18(4):e0284263, 2023.

\bibitem{rashidinia2018dynamical}
J.~Rashidinia, M.~Sajjadian, J.~Duarte, C.~Janu{\'a}rio, and N.~Martins.
\newblock On the dynamical complexity of a seasonally forced discrete sir epidemic model with a constant vaccination strategy.
\newblock {\em Complexity}, 2018(1):7191487, 2018.

\bibitem{read2021novel}
J.~M. Read, J.~R. Bridgen, D.~A. Cummings, A.~Ho, and C.~P. Jewell.
\newblock Novel coronavirus 2019-ncov (covid-19): early estimation of epidemiological parameters and epidemic size estimates.
\newblock {\em Philosophical Transactions of the Royal Society B}, 376(1829):20200265, 2021.

\bibitem{rodriguez2024machine}
A.~Rodr{\'\i}guez, H.~Kamarthi, P.~Agarwal, J.~Ho, M.~Patel, S.~Sapre, and B.~A. Prakash.
\newblock Machine learning for data-centric epidemic forecasting.
\newblock {\em Nature Machine Intelligence}, pages 1--10, 2024.

\bibitem{rohith2020dynamics}
G.~Rohith and K.~Devika.
\newblock Dynamics and control of covid-19 pandemic with nonlinear incidence rates.
\newblock {\em Nonlinear Dynamics}, 101(3):2013--2026, 2020.

\bibitem{schwartz1983infinite}
I.~B. Schwartz and H.~L. Smith.
\newblock Infinite subharmonic bifurcation in an seir epidemic model.
\newblock {\em Journal of mathematical biology}, 18:233--253, 1983.

\bibitem{setianto2023modeling}
S.~Setianto and D.~Hidayat.
\newblock Modeling the time-dependent transmission rate using gaussian pulses for analyzing the covid-19 outbreaks in the world.
\newblock {\em Scientific Reports}, 13(1):4466, 2023.

\bibitem{silva2022complex}
C.~J. Silva, G.~Cantin, C.~Cruz, R.~Fonseca-Pinto, R.~Passadouro, E.~S. Dos~Santos, and D.~F. Torres.
\newblock Complex network model for covid-19: human behavior, pseudo-periodic solutions and multiple epidemic waves.
\newblock {\em Journal of mathematical analysis and applications}, 514(2):125171, 2022.

\bibitem{soper1929interpretation}
H.~E. Soper.
\newblock The interpretation of periodicity in disease prevalence.
\newblock {\em Journal of the Royal Statistical Society}, 92(1):34--73, 1929.

\bibitem{statista_monthly_temperature_us}
{Statista}.
\newblock Monthly average temperature in the united states (in fahrenheit), 2023.
\newblock Accessed: 2023-09-01.

\bibitem{stolerman2015sir}
L.~M. Stolerman, D.~Coombs, and S.~Boatto.
\newblock Sir-network model and its application to dengue fever.
\newblock {\em SIAM Journal on Applied Mathematics}, 75(6):2581--2609, 2015.

\bibitem{stone2007seasonal}
L.~Stone, R.~Olinky, and A.~Huppert.
\newblock Seasonal dynamics of recurrent epidemics.
\newblock {\em Nature}, 446(7135):533--536, 2007.

\bibitem{suparit2018mathematical}
P.~Suparit, A.~Wiratsudakul, and C.~Modchang.
\newblock A mathematical model for zika virus transmission dynamics with a time-dependent mosquito biting rate.
\newblock {\em Theoretical Biology and Medical Modelling}, 15:1--11, 2018.

\bibitem{sutherst2004global}
R.~W. Sutherst.
\newblock Global change and human vulnerability to vector-borne diseases.
\newblock {\em Clinical microbiology reviews}, 17(1):136--173, 2004.

\bibitem{swei2020patterns}
A.~Swei, L.~I. Couper, L.~L. Coffey, D.~Kapan, and S.~Bennett.
\newblock Patterns, drivers, and challenges of vector-borne disease emergence.
\newblock {\em Vector-Borne and Zoonotic Diseases}, 20(3):159--170, 2020.

\bibitem{tabataba2017framework}
F.~S. Tabataba, P.~Chakraborty, N.~Ramakrishnan, S.~Venkatramanan, J.~Chen, B.~Lewis, and M.~Marathe.
\newblock A framework for evaluating epidemic forecasts.
\newblock {\em BMC infectious diseases}, 17:1--27, 2017.

\bibitem{tackney2023comparison}
M.~S. Tackney, T.~Morris, I.~White, C.~Leyrat, K.~Diaz-Ordaz, and E.~Williamson.
\newblock A comparison of covariate adjustment approaches under model misspecification in individually randomized trials.
\newblock {\em Trials}, 24(1):14, 2023.

\bibitem{tanaka2013effects}
G.~Tanaka and K.~Aihara.
\newblock Effects of seasonal variation patterns on recurrent outbreaks in epidemic models.
\newblock {\em Journal of theoretical biology}, 317:87--95, 2013.

\bibitem{trejos2022dynamics}
D.~Y. Trejos, J.~C. Valverde, and E.~Venturino.
\newblock Dynamics of infectious diseases: A review of the main biological aspects and their mathematical translation.
\newblock {\em Applied Mathematics and Nonlinear Sciences}, 7(1):1--26, 2022.

\bibitem{van2017reproduction}
P.~Van~den Driessche.
\newblock Reproduction numbers of infectious disease models.
\newblock {\em Infectious disease modelling}, 2(3):288--303, 2017.

\bibitem{wang2012simple}
R.-H. Wang, Z.~Jin, Q.-X. Liu, J.~van~de Koppel, and D.~Alonso.
\newblock A simple stochastic model with environmental transmission explains multi-year periodicity in outbreaks of avian flu.
\newblock {\em PLoS One}, 7(2):e28873, 2012.

\bibitem{wang2008threshold}
W.~Wang and X.-Q. Zhao.
\newblock Threshold dynamics for compartmental epidemic models in periodic environments.
\newblock {\em Journal of Dynamics and Differential Equations}, 20:699--717, 2008.

\bibitem{wang2017dynamics}
X.~Wang and X.-Q. Zhao.
\newblock Dynamics of a time-delayed lyme disease model with seasonality.
\newblock {\em SIAM Journal on Applied Dynamical Systems}, 16(2):853--881, 2017.

\bibitem{wiemken2020machine}
T.~L. Wiemken and R.~R. Kelley.
\newblock Machine learning in epidemiology and health outcomes research.
\newblock {\em Annu Rev Public Health}, 41(1):21--36, 2020.

\bibitem{yuan2021modeling}
H.~Yuan, S.~C. Kramer, E.~H. Lau, B.~J. Cowling, and W.~Yang.
\newblock Modeling influenza seasonality in the tropics and subtropics.
\newblock {\em PLoS computational biology}, 17(6):e1009050, 2021.

\bibitem{zhang2019dynamics}
J.~Zhang, Y.~Li, Z.~Jin, and H.~Zhu.
\newblock Dynamics analysis of an avian influenza a (h7n9) epidemic model with vaccination and seasonality.
\newblock {\em Complexity}, 2019(1):4161287, 2019.

\end{thebibliography}
\end{document}